\documentclass[manuscript,screen]{acmart}
\AtBeginDocument{%
  }

\setcopyright{acmlicensed}
\copyrightyear{2025}
\acmYear{2025}
\acmDOI{XXXXXXX.XXXXXXX}
\acmConference[Conference acronym 'XX]{Make sure to enter the correct
  conference title from your rights confirmation email}{June 03--05,
  2018}{Woodstock, NY}
\acmISBN{978-1-4503-XXXX-X/2018/06}

\usepackage[dvipsnames]{xcolor}

\usepackage{todonotes}

\usepackage{enumitem}
\usepackage{listings}
\usepackage{accsupp}
\begin{document}

\title[Invisible Impact of Empathy on Behavioral Change]{Invisible Impact of Empathy on Behavioral Change: Isolating the Effect of Empathy in Long-term Physical Activity Coaching Chatbot Interactions}

\author{Li Siyan}
\email{siyan.li@columbia.edu}
\orcid{0009-0002-1465-6337}
\affiliation{%
  \institution{Columbia University}
  \city{New York}
  \state{NY}
  \country{USA}
}

\author{Kai-Hui Liang}
\orcid{0000-0002-5567-848X}
\affiliation{%
  \institution{Columbia University}
  \city{New York}
  \state{NY}
  \country{USA}
}

\author{Shopnil Shahriar}
\orcid{0009-0005-9132-0206}
\affiliation{%
  \institution{Columbia University}
  \city{New York}
  \state{NY}
  \country{USA}
}

\author{Yilin Ye}
\orcid{0009-0008-8848-974X}
\affiliation{%
  \institution{Columbia University}
  \city{New York}
  \state{NY}
  \country{USA}
}

\author{Shiyoh Goetsu}
\orcid{0009-0002-6689-0512}
\affiliation{%
  \institution{Sony Group Corporation}
  \city{Tokyo}
  \country{Japan}
}

\author{Wei-Wei Du}
\orcid{0000-0002-0627-0314}
\affiliation{%
  \institution{Sony Group Corporation}
  \city{Tokyo}
  \country{Japan}
}

\author{Masahiro Yoshida}
\orcid{0009-0000-2394-1725}
\affiliation{%
  \institution{Sony Group Corporation}
  \city{Tokyo}
  \country{Japan}
}

\author{Tsunayuki Ohwa}
\orcid{0009-0005-7544-8648}
\affiliation{%
  \institution{Sony Group Corporation}
  \city{Tokyo}
  \country{Japan}
}

\author{Xuhai Xu}
\orcid{0000-0001-5930-3899}
\affiliation{%
  \institution{Columbia University}
  \city{New York}
  \state{NY}
  \country{USA}
}

\author{Zhou Yu}
\orcid{0000-0002-1524-5890}
\affiliation{%
  \institution{Columbia University}
  \city{New York}
  \state{NY}
  \country{USA}
}

\renewcommand{\shortauthors}{Siyan et al.}

\begin{abstract}
Current dialogue systems, powered by large language models, often treat empathy as essential without assessing its true impact, especially in behavior change, where motivation and adherence often depend on subtle user-chatbot dynamics. We examine this assumption by building three WhatsApp physical-activity (PA) coaching chatbots that differ only in empathy level and evaluating them in a six-week within-subject study (N = 13). Participants struggled to distinguish between the empathy conditions, and the non-empathetic version was often rated as more engaging and useful. However, higher-empathy variants were still associated with a larger overall average increase in step counts and faster improvement in intention to follow advice. These results suggest empathy’s role is nuanced: it may be hard for lay users to identify explicitly, but it can still shape motivation and trust that support sustained change. We interpret this pattern through the Elaboration Likelihood Model’s peripheral route. We highlight design implications for building next-generation PA coaching chatbots that balance effectiveness with human-like connection.

\end{abstract}

\ccsdesc[500]{Human-centered computing~Empirical studies in HCI}
\ccsdesc[300]{Human-centered computing~Natural language interfaces}
\keywords{Physical Activity Coaching, Behavioral Change, Chatbots, Empathy}

\keywords{Physical Activity Coaching, Behavioral Change, Chatbots, Empathy}


\maketitle

\section{Introduction}

Incorporating empathy \cite{clark1980empathy} into conversational systems has been a heavily studied topic \cite{rashkin2018towards,zhou2020design,liu2018should}. Empathy can enhance trust and long-term engagement \cite{bickmore2005establishing}, reduce depression symptoms in mental health support \cite{fitzpatrick2017delivering}, and facilitate the resilience process and enhance well-being \cite{jiang2022chatbot}. As a result, many applications of interactive systems emphasize empathy as a core design principle. One of these applications is Physical Activity (PA) coaching chatbots. 
These digital systems, developed by researchers to promote behavioral change \cite{piao2020use, kocielnik2018reflection}, demonstrate promising potential in improving physical activity and diet adherence \cite{singh2023systematic} and combat the inaccessibility of in-person PA counseling. 

\begin{figure}
    \centering
    \BeginAccSupp{
        method=pdfstringdef,
        Alt={Workflow of the empathetic Physical Activity chatbot. A clinical empathy module combines the system prompt, conversation history, and user utterance to generate an empathetic system instruction that populates the language model context and produces an empathetic response. The chatbot interacts with users via a mobile chat interface, illustrating how longer-term interactions can increase self-efficacy and intention to follow chatbot advice.}
    }
    \includegraphics[width=0.8\linewidth]{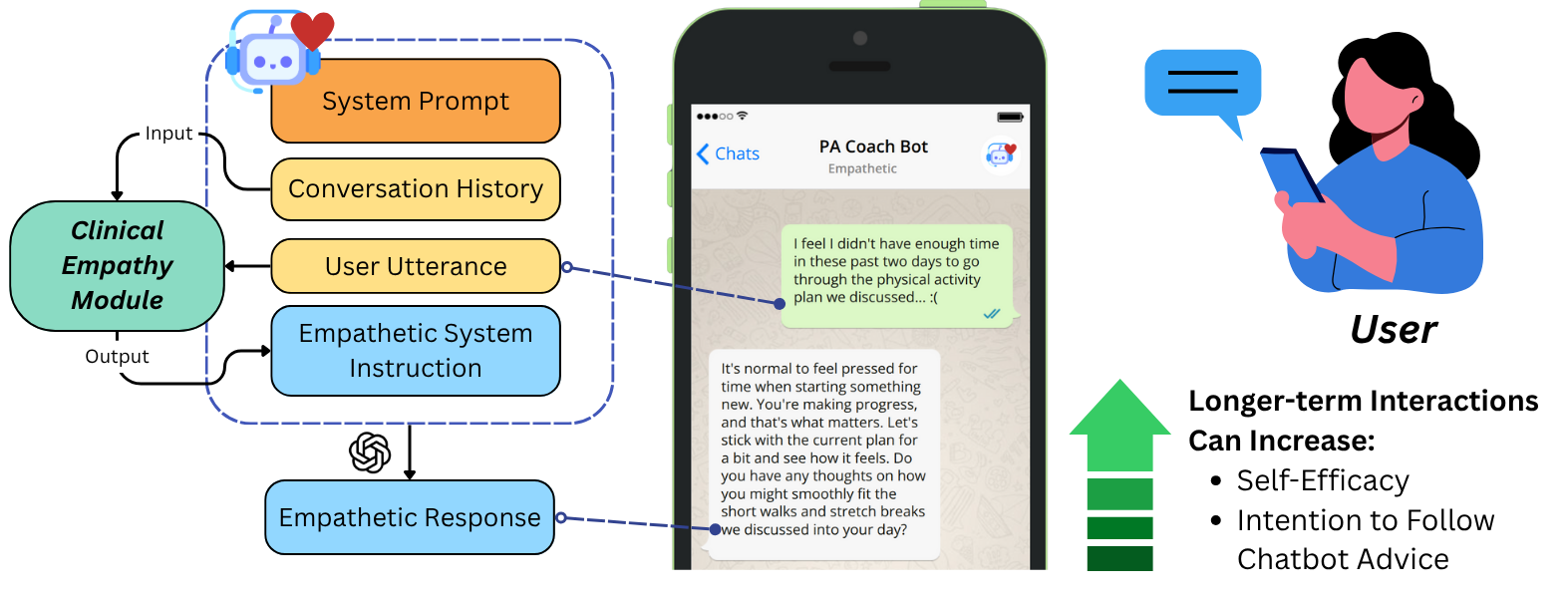}
    \EndAccSupp{}
    \caption{The overall workflow of the Empathetic version of our Physical Activity (PA) coaching chatbot.}
    \label{fig:placeholder}
\end{figure}

Most existing digital PA coaching systems utilize older natural language processing technologies. Nowadays, as LLMs shift the interaction paradigm, current research still focuses only on the initial consultation session in their design, neglecting sustained engagement and adherence \cite{jorke2024supporting}. Furthermore, many existing coaching systems utilize motivational interviewing techniques \cite{miller2002motivational}, often emphasizing empathetic responses, as guidelines for chatbot behaviors. From a theoretical perspective, higher degrees of empathy may enhance the social presence of the chatbot \cite{biocca2003toward}, thereby promoting self-motivation through improved relatedness (self-determination theory \cite{deci2013intrinsic}), which in turn drives sustained behavioral change. The effect of \textit{empathy alone as an isolated independent variable} has not been sufficiently studied in prior literature under longer-term digital PA counseling system settings. Understanding this effect is crucial because it can inform the design of more effective coaching systems, clarifying whether improvements in engagement and adherence rely on empathetic responses themselves or other counseling strategies. 

To isolate the effect of chatbot empathy on participant outcomes, we construct versions of a personalized, LLM-based physical activity coaching chatbot with varying degrees of empathy. Our chatbots engage with users through an initial consultation session and then a series of maintenance sessions. The initial session aims to acquaint the user with the chatbot, get to understand the user's situation, and provide tailored action plans. The maintenance sessions are designed to be more effective in engaging users through employing personalization, which has been shown to enhance engagement and satisfaction \cite{ha2024clochat}.
A total of three chatbot versions were designed for our study: \textbf{Non-Empathetic, Standard, and Empathetic}. The Non-Empathetic version is prompted not to respond empathetically; the Standard version leverages the emergent capabilities of LLMs to express empathy \cite{luo2024assessing}; the Empathetic version identifies and responds empathetically to Empathy Opportunities in user responses in maintenance sessions (all sessions after the initial consultation session). Here, the notion of Empathy Opportunities (EOs) refers to cues that the user would provide for others to respond empathetically, such as negative self-judgment (e.g. I could have done better) \cite{pounds2011empathy}. Upon identifying these opportunities, the Empathetic chatbot would respond empathetically in a manner similar to how healthcare professionals would respond to these messages, guided by study data from \citet{rey2024digital}. Our Empathetic chatbot therefore employs \textit{clinical} empathy, contrasting with the \textit{general} empathy embodied by the Standard chatbot.

We hypothesize that this theory-grounded mechanism for conveying empathy may strengthen the relationship between the user and the chatbot, making the users more engaged and more likely to heed the chatbot's suggestions. Specifically, receiving empathy from the chatbot may boost the user's self-efficacy and enhance the user's trust in the chatbot, resulting in overall increased PA levels and better engagement through improved rapport \cite{norfolk2007role} over multiple interactions. We propose the following research question:

\begin{quote}
    \textbf{RQ:} Does adopting a clinically-informed empathetic framework in a PA coaching chatbot positively influence user outcomes (including overall PA level and self-efficacy) and chatbot perception under a multi-session setting?
\end{quote}

To investigate this research question, we conducted a within-subject user study (N = 13) over six weeks. We measure self-efficacy, the user's relationship quality with the chatbot, and physical activity in terms of step counts, along with other metrics, using self-report questionnaires. 

Our results indicate that, even when participants often struggle to differentiate between the chatbot versions throughout the weeks, their behavior is still influenced by the presence of empathy. While the Non-Empathetic chatbot can be perceived as more engaging and useful than the Standard and Empathetic variants, the Empathetic chatbot increases participant intention to follow chatbot advice and self-efficacy significantly faster, and is associated with higher trustworthiness compared to the Non-Empathetic version. Additionally, initiating the six-week interaction with the Empathetic chatbot results in the least attrition. This work underscores the nuanced and positive effect of empathy on behavioral change. The contribution of this work is as follows:
\begin{enumerate}
    \item We incorporate a clinical empathy framework informed by \citet{rey2024digital} and \citet{pounds2011empathy} into a PA coaching chatbot.
    \item We systematically examine the isolated effect of empathy using three PA chatbot variants using a within-subject, six-week-long user study.
    \item We reveal that focusing solely on reported perceived empathy from non-expert participants for chatbot assessment may not be a reliable method.
    \item We present design considerations for future LLM-powered PA coaching chatbots.
\end{enumerate}

\section{Related Work}

\subsection{Chatbots for Physical Activity Coaching}

Prior to modern-day transformer-based LLMs, chatbots designed for Physical Activity (PA) coaching relied on fixed and templated responses and workflows through intent recognition \cite{larbi2022usability, to2021feasibility}, which can be overly rigid and generalized. Process evaluations and component analyses further highlight engagement challenges, such as participant frustration from communication failures \cite{larbi2022usability, davis2020process} and attrition \cite{kramer2020components}. Even with these limitations, systematic reviews on these pre-LLM chatbots reveal that they are still efficacious in elevating PA levels, with an average effect of +735 steps a day across 19 trials \cite{singh2023systematic}. It is plausible that LLM-powered chatbots, now with potential for higher flexibility and personalized support \cite{zhang2024personalization}, would be able to better enhance the positive outcomes from interacting with these chatbots. 

More recently, PA coaching chatbot designs have been incorporating modern LLMs to varying degrees. \citet{vandelanotte2023increasing} discusses the implementation of MoveMentor, a chatbot capable of providing just-in-time interventions. Despite utilizing LLMs to generate question-answer pairs relevant to their use case, the authors do not directly leverage LLMs for handling all interactions, but instead an intent-based retriever. This prevents MoveMentor from being able to utilize the full range of LLM capabilities, especially since LLM generations may lack overall diversity \cite{wang2025multilingual}. 

One of the first fully LLM-based PA chatbots is GPTCoach \cite{jorke2025gptcoach}. This system implements the onboarding session using Motivational Interviewing (MI), a core empathetic technique leveraged for behavioral change coaching \cite{miller2002motivational, bischof2021motivational}. GPTCoach additionally provides visualizations of historical PA data to enrich its advice. The authors highlight the need to cater LLM-based coaching chatbot advice to each individual participant, which we address in our design through personalized action planning. A limitation of this work is that GPTCoach only supports the initial consultation session, leaving the longer-term efficacy of LLM-based chatbots underexplored.

Providing data-driven guidelines for designing PA coaching chatbots, \citet{liang2021evaluation} examines a set of human PA counseling dialogues. The authors categorized utterances into counseling strategies and topics. They discover that barriers and motivation are heavily discussed during these counseling sessions, and that more barrier-relevant strategies from the counselor is correlated with improvements in the barrier survey.

Generally, we see that most chatbots for positive behavioral change are yet to embrace state-of-the-art NLP technologies. There is additionally a lack of longer-term evaluations for such chatbots employing modern LLMs. Our work bridges this important gap by studying the longer-term impact of an LLM-based physical activity coaching chatbot.

\subsection{Role of Empathy in Behavioral Change}

Empathy has been a robust relational factor in counseling. Therapist empathy exhibits a moderately strong correlation with better client outcomes \cite{elliott2018therapist}. In the behavioral change context, Motivational Interviewing (MI) is a common approach that underlines empathy as its core principle. MI has been shown to significantly increase the self-efficacy of type 2 diabetes patients \cite{hemmati2021effect} and promote weight loss and self-efficacy improvement among adolescents \cite{walpole2011motivational}. 

In digital counseling systems, empathy is frequently a design principle but remains under-isolated as a causal factor. Existing empathetic conversational agents for behavioral change employ rule- and template-based approaches \cite{mercado2023embodied}, often constrained to multiple-choice input, and have short study durations and limited personalization \cite{luo2021promoting}. 

Early empathetic embodied agents are perceived positively but rarely tested for long-term effects on motivation and self-efficacy. ODVIC is a seminal work in incorporating explicit verbal and facial empathy while delivering brief MI-style interventions for excessive drinking \cite{lisetti2013can}. Participants perceived the empathetic agent to be more human-like and exhibited higher levels of acceptance, yet outcomes such as sustained motivation or self-efficacy were not evaluated longitudinally. Relational-agent work similarly demonstrates that social-emotional dialog and empathic behaviors can increase alliance and short-term PA (e.g., increased walking among older adults) \cite{bickmore2005establishing,bickmore2013randomized}. \citet{rahmanti2022slimme} proposes SlimeMe, a weight management chatbot equipped with artificial empathy through text-based emotion analysis coupled with emojis and other emotional chatbot responses, implemented with the intention-based DialogFlow framework. Participants perceived SlimeMe to be exciting and engaging after a 7-day simulation trial. Recent counseling systems begin to integrate LLMs and to target empathy in a more flexible manner, but they are still in an early stage. An MI chatbot with generative reflections produced larger short-term improvement in readiness and confidence for smoking cessation compared to its non-generative counterpart \cite{brown2023motivational}. Still, longer study durations and rigorous testing are still required to examine the effectiveness of empathy in behavioral change coaching systems.

Closer to physical activity and empathy, \citet{chauvin2023virtual} closely examines the effect of empathy on physical activity among older French adults. The authors devise a set of three interventions, two with non-directive MI interventions and one that directs the conversation using MI principles. The non-directive MI virtual agent is perceived as more empathetic and successfully improves participants' barrier self-efficacy and intrinsic motivation. However, participants only received a single session, outcomes were measured one week later, and all variants embodied MI-consistent empathy. Therefore, to further determine the effect of explicit empathy on self-efficacy, there needs to be longer-term user studies with more frequent interventions.

\section{Design Principles}

Prior work largely reports high-level guidelines and prompts but rarely specifies utterance-level behavior or provides open-source implementations. As a result, our implementation of the physical coaching chatbot is constructed from scratch, rather than extending an existing PA chatbot implementation. We derive our design principles based on the behavioral change literature.

\begin{itemize}
    \item \textbf{Principle 1. Theory-Grounded Implementation:} Identifying and implementing theory-grounded interventions strengthens the construct validity of our design. We will utilize theoretical frameworks such as motivational interviewing when implementing the chatbot, similar to prior PA chatbot implementations \cite{jorke2024supporting,chauvin2023virtual}. Additionally, current implementations of \textit{empathetic} conversational systems for behavioral change are not sufficiently grounded in operationalized empathy in the physical activity coaching context. We will bridge this gap by employing a clinical empathy framework from \citet{pounds2011empathy}.
    \item \textbf{Principle 2. Design in accordance with Human PA Counseling:} Human PA counseling and therapy have been more extensively studied than their digital counterparts. Due to the increasing anthropomorphism toward LLMs such as ChatGPT \cite{jacobs2023brief,polyportis2025understanding}, it is possible that transplanting successful human PA counseling strategies to chatbot PA counseling scenarios would yield successful results. Using inspection results from \citet{liang2021evaluation}, we will place emphasis on addressing barriers and motivation in our initial session. In maintenance sessions, we will aim to incorporate chatbot questions that elicit discussion of barriers and motivation.
    \item \textbf{Principle 3. Better Personalization for Enhanced Long-Term Engagement:} Prior work has indicated that the rigidity of pre-LLM chatbots suffers from attrition issues. As our work aims to study longer-term effects, sustained engagement from the users is desirable, which can be addressed through higher application flexibility (already handled by LLM emergent capabilities) and through better personalization \cite{ha2024clochat}.
    \item \textbf{Principle 4. Experimentally Controllable Empathy:} We will hold the core content of the chatbot consistent across the three conditions, such that the primary difference between them will be stylistic differences in exhibited and perceived empathy. This can be achieved through consistent system prompts throughout different chatbot versions.
\end{itemize}

\section{Chatbot System Design}

In this section, we detail the construction of our physical activity coach chatbot. 
All chatbot variants use the same base model (GPT-4o). We chose a general-purpose LLM rather than specialized fine-tuned models because public physical-activity coaching dialogues are limited, keeping the base model constant controls for capability differences across conditions, and our study requires multiple variants. This also enables rapid iteration while reducing the risk that fine-tuning would introduce confounds beyond the targeted empathic features. Our Empathetic chatbot features a clinical empathy module; the other two versions do not have access to this module. Instead, we instruct GPT-4o not to be empathetic when responding as the Non-Empathetic chatbot; for the Standard chatbot, we leverage the same system instructions as the Empathetic version, but without the clinical empathy modules, relying on the innate LLM capabilities to express empathy.


Since personalization has been shown to facilitate long-term engagement \cite{ha2024clochat}, the three chatbot versions also incorporate personalization. Personalization is realized in multiple ways in our chatbot, after considering approaches to benefiting PA plan adherence. First, our chatbot provides personalized action planning tailored to the user's specific lifestyle and living situations, rendering the PA action plans more feasible. Secondly, our chatbot provides stress relief to users using short micro-interventions \cite{johnson2022efficacy,paredes2014poptherapy} for users who are experiencing excess stress, as stress reduction has been shown to increase PA \cite{stults2014effects}. Thirdly, our chatbot retains past conversations in a condensed format and retrieves the last three conversations to propose a pertinent discussion topic during maintenance sessions, to center the conversation on the user.

We implement our chatbot as a WhatsApp bot with Twilio integrations. Reminder messages for engaging with the WhatsApp chatbot are sent through SMS texts every 48 hours. We had to resort to this less integrated design choice because Twilio no longer supports scheduled WhatsApp messages more than 24 hours in advance.

\begin{figure}
    \centering
    \BeginAccSupp{method=pdfstringdef,
        Alt={Comparison diagram showing the context provided to the language model during maintenance sessions across three chatbot versions. The empathetic version includes a clinical empathy module that combines system prompt, conversation history, user utterance, and empathetic system instruction. The standard version consists of only the system prompt, conversation history, and user utterance. The non-empathetic version replaces empathy-related instructions with non-empathetic system instructions, illustrating differences in how empathy is incorporated into the model context.}
    }
    \includegraphics[width=0.9\linewidth]{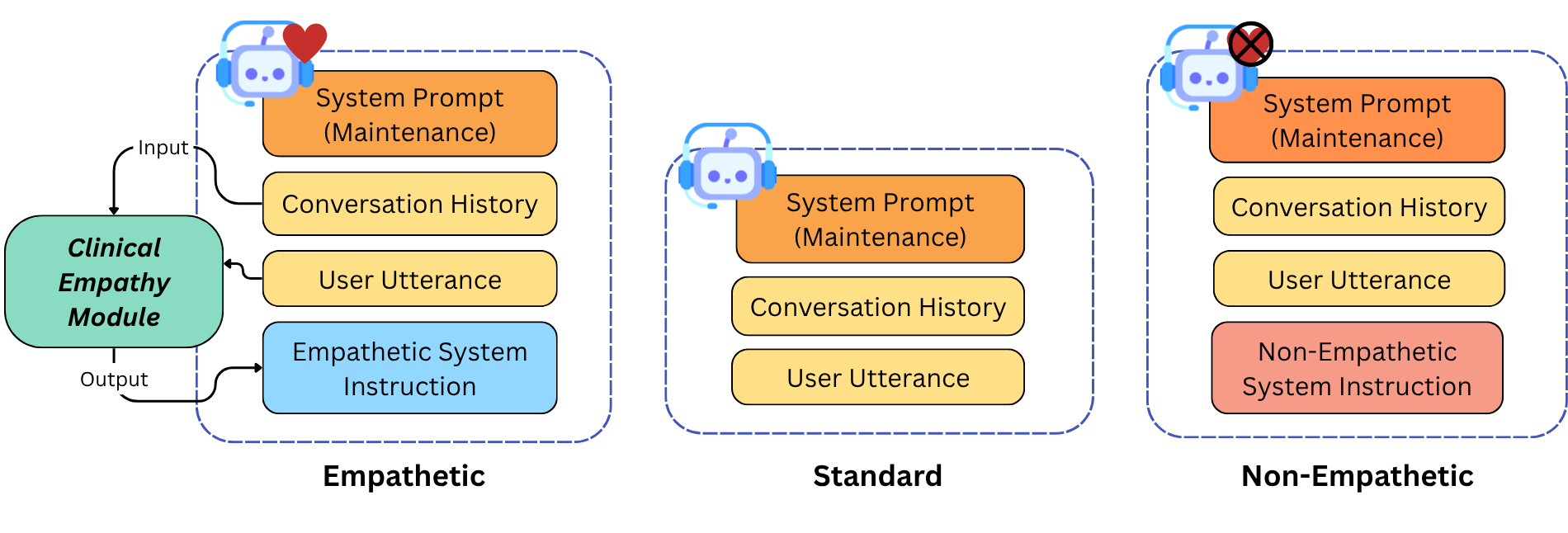}
    \EndAccSupp{}
    \caption{Comparison of what the context fed into the LLM would look like between the three versions of the chatbot in the maintenance session.}
    \label{fig:overall_flow_comparison}
\end{figure}

\subsection{Designing the Initial Session}
In the initial session, the chatbot follows the Motivational Interviewing procedur \cite{miller2002motivational}. The Empathetic and Standard chatbot versions share the prompts for this session, and we do not activate the clinical empathy modules in this session. The Non-Empathetic chatbot uses a gently rephrased system prompt that removes the requirement for empathy in the responses.

Following Principle 2, we address barriers and motivation directly in our initial session. The guiding principles for the four stages of the initial session are as follows:
\begin{enumerate}
    \item \textit{Engaging:} Getting to know the user, forming rapport.
    \item \textit{Focusing:} Identifying barriers, exploring possible solutions, tailoring suggestions to the user's personal situation and preferences.
    \item \textit{Evoking:} Reinforcing and enhancing motivation.
    \item \textit{Planning:} Decide and discuss a feasible short-term action plan.
\end{enumerate}

See the full system prompts and an example conversation in Appendix \ref{app:prompts_and_ex}. For both the initial and maintenance sessions, we prompt the chatbot with specific talking points to better control empathy as the independent variable. As conversations in our study are short and we do not process external data or utilize tool calls, as in \citet{jorke2024supporting}, we are more confident that our chatbot can follow instructions without additional prompting techniques. 

\begin{figure}
    \centering
    \BeginAccSupp{method=pdfstringdef,
        Alt={Process diagram of the maintenance session showing how the chatbot conducts check-ins with users. The session begins with a ping message informed by a summary of the previous session, followed by questions about the user's current state, recent physical activity, and emotional well-being. Information from the previous three sessions is used to select a personalized topic and assess the user's stress level on a 0-5 numeric scale. The scale is then used to determine whether a stress relief micro-intervention is used.}
    }
    \includegraphics[width=0.8\linewidth]{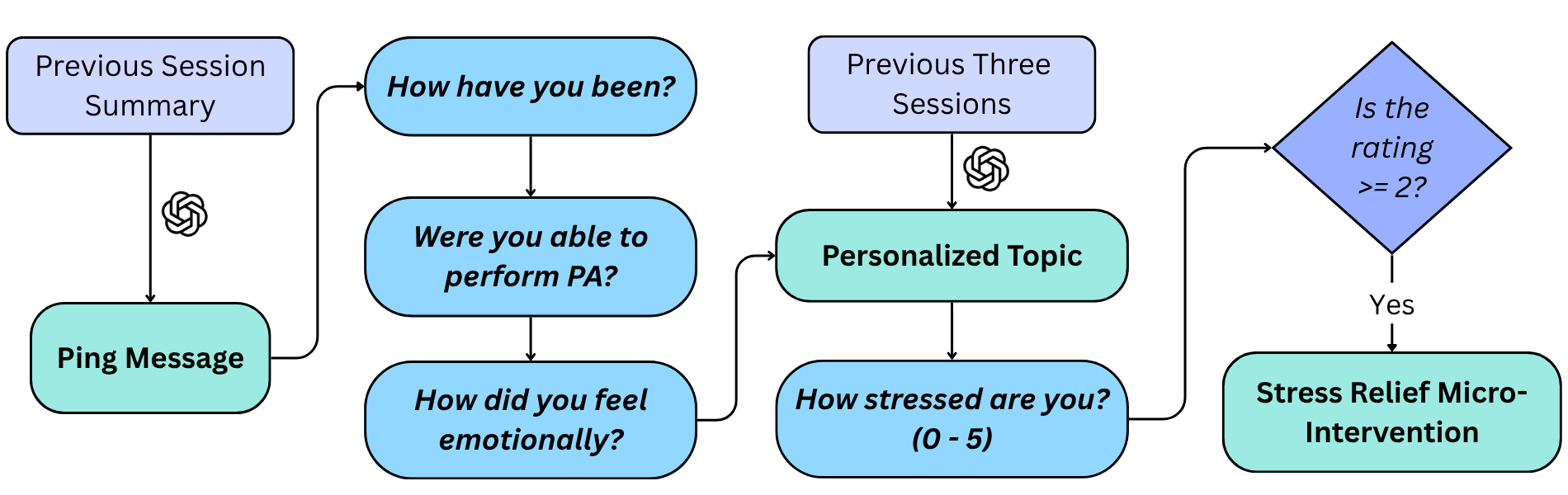}
    \EndAccSupp{}
    \caption{The general processes for the maintenance session. The personalized components are highlighted in green.}
    \label{fig:processes_convo}
\end{figure}

\subsection{Designing the Maintenance Sessions}
\begin{figure}
    \centering
    \BeginAccSupp{method=pdfstringdef,
        Alt={Example maintenance session conversations from the same user comparing empathetic, standard, and non-empathetic chatbot responses before the stress relief module. The empathetic chatbot acknowledges the user's feelings, validates their experience, and tailors the response with supportive language. The standard chatbot focuses on adjustments with limited emotional acknowledgement. The non-empathetic chatbot responds with no emotional acknowledgement and is focused on next steps and action planning.}
    }
    \includegraphics[width=\linewidth]{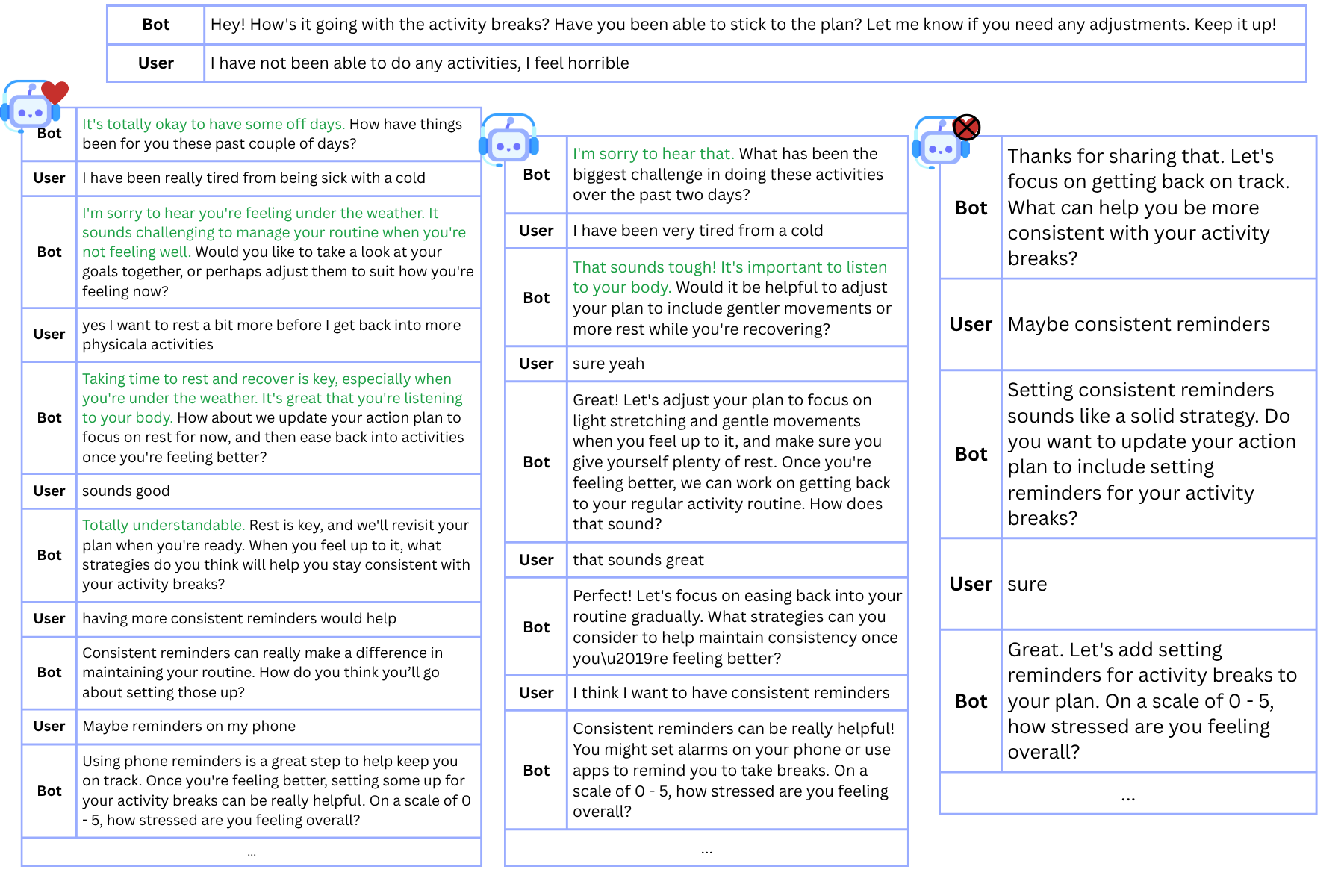}
    \EndAccSupp{}
    \caption{Example maintenance session conversations from the same user with Empathetic (left), Standard (middle), and Non-Empathetic (right) chatbots, up until the stress relief workflow. We highlight explicitly empathetic expressions in green.}
    \label{fig:ex_convo}
\end{figure}

Figure \ref{fig:overall_flow_comparison} indicates the differences between the three chatbot versions when processing incoming user messages during the maintenance session. The general maintenance workflow is held consistent among all chatbot versions. 
We include example conversations for each of the chatbots in Figure \ref{fig:ex_convo}.

After receiving the SMS reminder message, the user would first send a pre-defined message to trigger an LLM-driven summarization process, which condenses the previous conversation into user-centric bullet points. Using these, GPT-4o would generate a personalized ping message.
A potential topic to discuss, as well as any additional relevant context, would then be incorporated into the chatbot's context window (Section \ref{sec:long_term_mem}). The chatbot would finally transition into asking the user a set of check-in questions.

When designing these check-in questions, we aim to create questions that may elicit negative self-disclosure. This is to create more naturalistic Empathy Opportunities for the Clinical Empathy modules (Section \ref{sec:clinical_empathy}). We arrived at the same three questions for each maintenance session:

\begin{enumerate}
    \item How have you been for the last two days?
    \item Were you able to perform the physical activity previously discussed?
    \item How were you feeling emotionally when you were doing the physical activities?
\end{enumerate}

Finally, the chatbot would ask the user to rate their stress level on a scale of 0-5 and provide stress-relief micro-interventions if the stress level is 2 or higher (Section \ref{sec:stress_relief}).

\subsubsection{Long-term Memory Module}
\label{sec:long_term_mem}

The Long-term Memory module maintains all bullet-point summaries, including the dates, of previous conversation sessions. Summaries are embedded using Sentence Transformers \cite{reimers-2019-sentence-bert} and stored in a FAISS index \cite{douze2024faiss} for easy retrieval.

Prior to sending the initial ping message in every maintenance session, we would prompt GPT-4o to determine a suitable discussion topic for the chatbot for this upcoming session, using the bullet point summaries of the previous three sessions. Using this suggested topic, we would then retrieve the top 3 - 5 relevant summary bullet points.

A limitation in this approach is that the chatbot would not retain \textit{every single} conversation. We originally posit that this should not be a significant issue, because questions in the maintenance session are highly focused on the recent past, and the customized discussion topic should already have sufficient background information through retrieval for informative discussion. We are further incentivized to keep the context shorter due to the possible degradation in instruction-following that can result from long contexts in LLMs.

\begin{figure}
    \centering
    \BeginAccSupp{method=pdfstringdef,
        Alt={Workflow diagram of the Clinical Empathy Module used to populate the language model context with the appropriate empathy system instruction. Conversation history and the current user utterance are inputs into Empathy Opportunity classifier, which predicts multiple empathy opportunities present in the user's message. The predictions are mapped to specific empathy strategies derived from prior healthcare communication data, which are combined and sampled by a strategy sampler. The selected response strategy is passed to an instruction generator that produces empathetic system instructions.}
    }
    \includegraphics[width=0.8\linewidth]{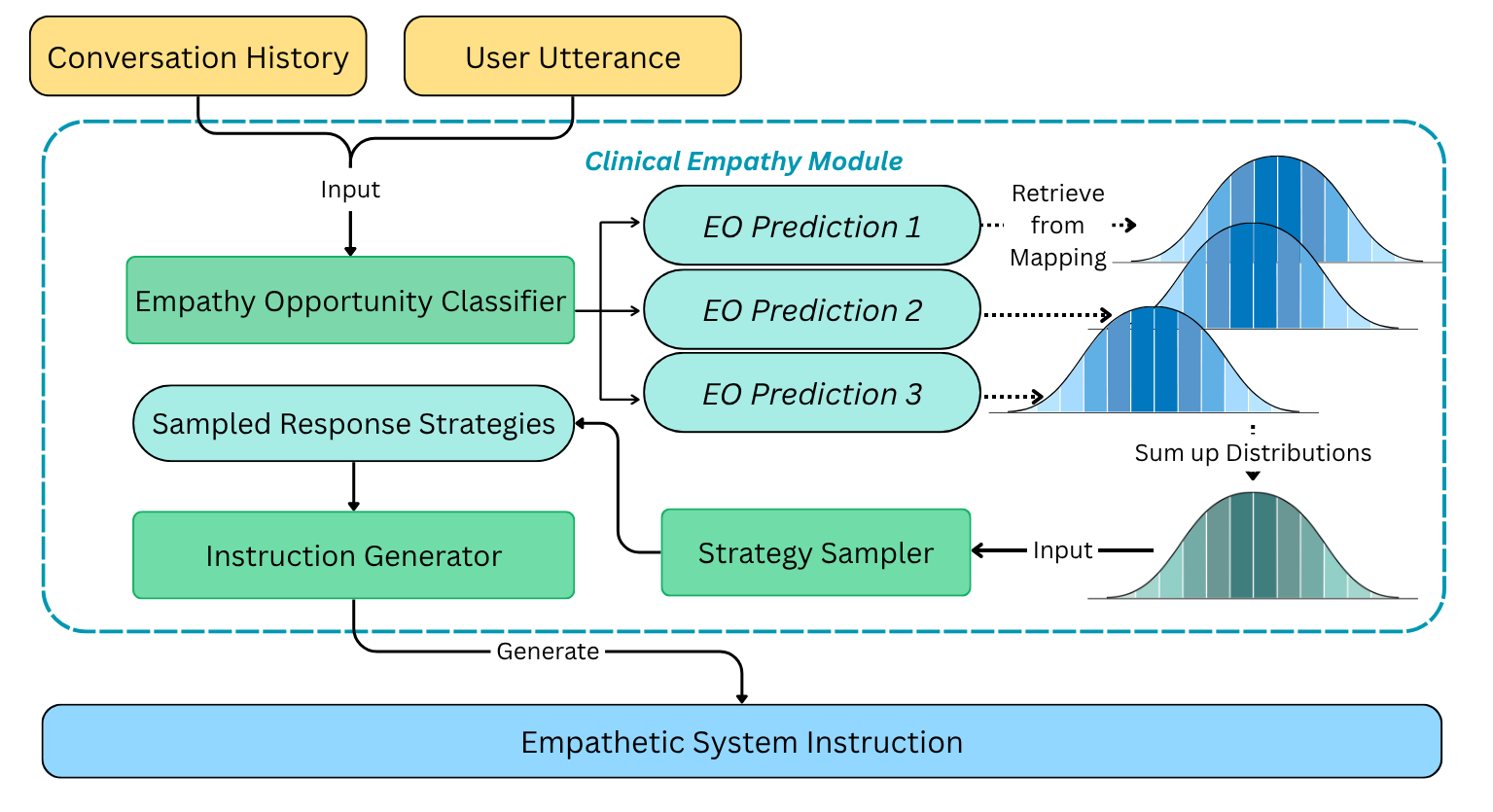}
    \EndAccSupp{}
    \caption{The overall workflow of the Clinical Empathy Module to populate the language model context with the appropriate empathy system instructions.}
    \label{fig:empathy_workflow}
\end{figure}

\subsubsection{Clinical Empathy Module}
\label{sec:clinical_empathy}


The definition of our chatbot's empathy framework is informed by a digital health study \cite{rey2024digital} investigating how healthcare professionals respond to messages from patients in a healthy lifestyle intervention program.
The authors first translate the definitions of Empathy Opportunities (EOs) and response strategies documented in \citet{pounds2011empathy} into the context of behavioral change, then annotate sentences from patient messages for the EOs they exhibit at a phrasal granularity. Afterwards, healthcare professionals responded to individual sentences presented in the messages.
Finally, the authors mapped the healthcare professionals' response strategies back to the response strategies from \citet{pounds2011empathy}.
This mapping forms the backbone of how our Empathetic chatbot responds to participants.

To integrate this framework into our chatbot design, we created three submodules: Empathy Opportunity Classifier, Strategy Sampler, and Instruction Generator. See an overview of this workflow in Figure \ref{fig:empathy_workflow}. We discuss the specific technical instantiation of our module in further detail in Appendix \ref{app:empathy_tech_details}.

\noindent \textbf{Empathy Opportunity Classifier:} Given a user message, we would first need to identify the Empathy Opportunities (EOs) present. We optimized a GPT-4o-mini-based classifier that achieves a top-three classification accuracy of 85.3\%. 

\noindent \textbf{Strategy Sampler:} We created EO-specific strategy distributions using mappings between EOs and empathetic response strategies using statistics from  \citet{rey2024digital} (e.g., when the user explicitly exhibits negative feelings, the healthcare professional responds with neutral support and explicit appreciation 37.5\% of the time). These distributions are then combined and sampled from, given the top-three EO predictions from the EO Classifier.

\noindent \textbf{Instruction Generator:} Finally, we create an instruction for GPT-4o to utilize descriptions of the selected empathetic response strategies. The instruction contains descriptions of the chosen strategies (modified from \citet{pounds2011empathy}).

\subsubsection{Stress Relief Module}
\label{sec:stress_relief}
The link between PA and stress is a reciprocal one: PA reduces stress, while more stress often predicts decreased physical activity \cite{stults2014effects}. As a means to provide more varied content to keep users better engaged, and to help them achieve their physical activity goals, we design the following stress relief module to provide in-conversation stress-relief micro-interventions. These micro-interventions follow the format of Prompt + URL \cite{paredes2014poptherapy}, and research indicates that even brief interventions can be effective in significantly reducing perceived stress \cite{yehuda2024improving}. Methodologically, just-in-time adaptive interventions and microrandomized trials provide frameworks for context-sensitive delivery and causal optimization in mobile settings \cite{NahumShani2018JITAI,Klasnja2015MRT,Klasnja2019HeartSteps}. 

We manually curate 23 micro-interventions spanning four primary categories: relaxation, response modulation (e.g., cognitive change), positive experiences, and resource buffers \cite{johnson2022efficacy}. See high-level descriptions of the categories in Appendix \ref{app:desc_stress_relief}. 
Similar to prior work \cite{ameko2020offline}, we employ a multi-armed bandit approach to select the optimal micro-intervention for the current user, and then utilize retrieval-augmented generation to generate the response. 

At the end of the maintenance sessions, the chatbot would prompt the user to provide a rating of their perceived stress on a scale of 0 - 5. If the current perceived stress is higher than 2, a micro-intervention will be administered through the multi-armed bandit. Afterwards, the user would again be asked to rate the effectiveness of the intervention on a scale of 1 through 10, which would be used to update the weights for the bandit.









\section{Pilot Study}

We conducted a between-subject pilot study with 21 participants through Prolific to explore whether different versions of the chatbot have short-term effects on self-efficacy in participants, while also ensuring the workflow functions as intended. There were eight participants in the Non-Empathetic condition, four in the Standard condition, and nine in the Empathetic condition. The distribution of participants was uneven because our primary interest was contrasting between the Empathetic and Non-Empathetic conditions. See more details about this pilot study in Appendix \ref{app:pilot_study}. 

\subsection{Study Design}

The study simulates a user's first two interactions with our coaching chatbot. After completing a pre-survey, participants undergo an initial chatbot session, followed by a maintenance session a few minutes later, simulating that two days have passed and they have not met their physical activity goals. The study concludes with a post-survey:

\begin{itemize}
    \item System Usability Scale \cite{brooke1996sus} with the word ``Computer System'' replaced with ``Chatbot''.
    \item \textbf{Chatbot Conversation Quality:} These questions are used to assess whether the chatbot responses have sufficiently high quality.
    \begin{enumerate}
        \item Did the chatbot provide naturalistic responses? (1 - The chatbot did not provide any naturalistic response; 5 - The chatbot always provided naturalistic responses)
        \item How can the chatbot be improved? Did anything feel odd or confusing during the conversation?
        \item What part of the conversation did you like the most?
        \item What part of the conversation did you dislike?
    \end{enumerate}
    \item \textbf{Conversation Outcome:} These Likert-style questions are reused for the check-in forms during our official user study. The questions are on a scale from Strongly Disagree to Strongly Agree.
    \begin{enumerate}
        \item \textbf{Intention to follow:} I intend to follow the chatbot's recommendation over the next 7 days.
        \item \textbf{Self-efficacy:} If I do intend to follow the chatbot's recommendation, I am confident that I can follow the recommendation over the next 7 days.
        \item \textbf{Recommendation Helpfulness:} I think the chatbot's recommendation is useful for enhancing my physical activity.
    \end{enumerate}
\end{itemize}

\subsection{Results}


Surprisingly, the Non-Empathetic chatbot was rated more favorably on several usability dimensions. However, due to the small sample size, we present these results descriptively (means/medians) in Table \ref{tab:usability_pilot} and do not consider them conclusive. The Non-Empathetic chatbot was generally rated as more enjoyable and satisfying, while the Empathetic chatbot was rated as easier to use.

Regarding the most enjoyable aspects of their sessions, multiple participants highlighted the stress-relief component. Many also appreciated personalized advice, suggesting that these modules enhance engagement.

We notice that chatbot perception is influenced by significant individual variations. All three versions of the chatbot can be perceived as friendly and even empathetic according to participant free-form feedback. However, participants place emphasis differently when describing these versions. Both the Empathetic and the Standard chatbots received the descriptor ``like chatting with a friend", while the Empathetic chatbot is praised for its positive energy (``it genuinely cares about your well-being'').

\subsection{Pilot Study Takeaways}

The pilot study challenges a common assumption that users always prefer empathetic coaching chatbots. Findings suggest people vary in detecting empathy and have different preferences for empathy and personalization. Based on this, we added pre-survey questions about desired empathy and personalization levels for the main study.

The study further highlights how individual differences can affect results. To address this, we chose a within-subject design to better compare the impact of different empathetic conditions.

\section{Longitudinal User Study}

To examine our hypotheses about the effect of empathy on user engagement, self-efficacy, and physical activity when interacting with a physical activity coach chatbot, we conducted a within-subject, six-week user study. Our study is approved by our home institution's IRB.

\subsection{Experimental Design}

After providing their informed consent, participants fill out a pre-survey:
\begin{enumerate}
    \item Current physical activity level, as measured by daily step counts and by the Stanford Brief Activity Survey \cite{taylor2010validation}.
    \item Stage of behavioral change. The specific question is as follows:\\
    \textbf{Question:} Please select the one that describes your current physical activity status.
\begin{enumerate}[label=(\Alph*)]
    \item I am currently not physically active and do not intend to engage in physical activity in the next six months
    \item I am currently not physically active, but I am thinking about getting more physically active in the next six months
    \item I currently do some physical activity, but not regularly
    \item I am currently physically active, but have only begun doing so within the last six months
    \item I am currently physically active and have done so for more than six months
\end{enumerate}
    \item Perceived Stress Scale \cite{cohen1983global}.
    \item Chatbot behavior preferences: 
    \begin{itemize}
        \item \textit{Do you prefer the coaching chatbot to be more empathetic or more direct (while not necessarily being as empathetic)?} \textbf{Ratings:} 1 (More direct) to 5 (More empathetic)
        \item \textit{Do you prefer having more generic suggestions from the chatbot (so that the chatbot does not need to know your personal information) or more personalized suggestions?} \textbf{Ratings:} 1 (More generally applicable suggestions) to 5 (More personalized suggestions)
    \end{itemize}
     Should the chatbot provide more personalized responses?
\end{enumerate}

Upon completing the pre-survey, participants would then proceed to WhatsApp to converse with the chatbot. The first session is the initial session, during which the chatbot gathers the user's information. All subsequent sessions, occurring every 48 hours, would be maintenance sessions. Each version of the chatbot (Non-Empathetic, Standard, Empathetic) is presented to every participant in a \textbf{partially counter-balanced} design. We did not use a fully counterbalanced design due to sample size constraints. Specifically, we leverage these experimental conditions:

\begin{enumerate}
    \item \textbf{Condition 1:} Non-Empathetic (NE), Standard (S), Empathetic (E), NE, S, E
    \item \textbf{Condition 2:} S, E, NE, S, E, NE
    \item \textbf{Condition 3:} E, NE, S, E, NE, S
\end{enumerate}


After each conversation, participants fill out a short Google Form, which only collects the Participant ID as identification. See further privacy considerations in Appendix \ref{app:privacy_protection}. The form contains the following:

\begin{enumerate}
    \item \textbf{Step Count:} What has been your average step count for the past two days?
    \item \textbf{Intention to follow:} I intend to follow the chatbot's recommendation over the next 7 days.
    \item \textbf{Self-efficacy:} If I do intend to follow the chatbot's recommendation, I am confident that I can follow the recommendation over the next 7 days.
    \item \textbf{Chatbot Usefulness:} I think the chatbot's recommendation is useful for enhancing my physical activity.
\end{enumerate}

For the check-in form, we aimed for a low-friction experience and therefore resorted to customized measures rather than longer standardized ones. Future work reproducing this study should consider leveraging standardized measures for better robustness.

We emphasized that participants have a means to track their steps in our instructional emails, aiming to improve the accuracy of self-reported step count tracking.

Additionally, participants complete a weekly reflection form on their experience from the previous week:

\begin{enumerate}
    \item Multi-Dimensional Measure of Trust \cite{ullman2019mdmt}, shortened.
    \item User Engagement Scale - Short Form \cite{o2018practical}, shortened.
    \item System Usability Scale with the word ``Computer System'' replaced with ``Chatbot''.
    \item Freeform text feedback (optional).
\end{enumerate}


\subsection{Participant Recruitment}

We recruited participants through social media and through emailing computer science students at the authors' home institution. 
In this study, we focus on inactive but motivated participants (those answering B-C to Question 2 in the pre-survey about their stage of behavior change).
106 users entered the pre-survey, and we narrowed it down to 33 after filtering.
Participant compensation is \$60; dropped-out participants are compensated proportional to the duration they were enrolled.


Our study began with 33 users. 16 of them dropped out before interacting with all three versions of the chatbot, 3 did not fill out surveys consistently, and 1 was excluded due to incorrect initial setup. Our analysis focused on the remaining 13 users (10 Females, 3 Males, aged 19-30) with high-quality data. 
More analysis about our attrition issue in Section \ref{sec:attrition}.

Our participants' demographics are listed in Table \ref{tab:participant_info}. We include additional information about our participants' familiarity with LLM chatbots and their attitudes toward Generative AI in Appendix \ref{app:llm_prefs_participants}. We see that the majority of our participants are women and young age groups, which we acknowledge as a limitation in our user study.
While participants have more variation in preferences for empathy, we observe that most participants share a positive attitude toward personalization.

\begin{table}[]
    \centering
    \begin{tabular}{cc|cc|cccc}
    \toprule
        \textbf{Condition} & \textbf{Exp. ID} & \textbf{Age} & \textbf{Gender} & \textbf{Step Count} & \textbf{Stress Level} & \textbf{Emp. Preference} & \textbf{Per. Preference} \\
        \midrule
            0 & P1& 22 & F & 700 & Moderate & 1 & 2 \\
            0 & P2& 19 & F & 3500 & Moderate & 2 & 4 \\
            0 & P3& 28 & M & N/A & High & 1 & 4 \\
            0 & P4& 23 & F & 4000 & Moderate & 3 & 4 \\
            0 & P5& 23 & F & 3500 & High & 3 & 4 \\
            1 & P6& 29 & F & 100 & High & 3 & 5 \\
            1 & P7& 32 & F & 500 & Moderate & 4 & 4 \\
            1 & P8& 24 & F & 5400 & High & 2 & 5 \\
            2 & P9& 25 & F & N/A & Moderate & 1 & 4 \\
            2 & P10& 20 & M & 6000 & Low & 1 & 5 \\
            2 & P11& 20 & M & 2000 & Low & 3 & 3 \\
            2 & P12& 22 & F & 1500 & Moderate & 4 & 4 \\
            2 & P13& 30 & F & 4000 & Moderate & 3 & 5 \\
    \bottomrule
    \end{tabular}
    \caption{Participant demographics, physical activity, stress, and PA chatbot preferences information from our user study.}
    \label{tab:participant_info}
\end{table}

\subsection{Result Analysis}
\label{sec:results_longitudinal}

In total, we collected 239 complete conversations. In this section, we present patterns identified through analyzing responses to the results from our recruitment pre-survey, post-conversation check-in, and weekly self-reflection forms.

\subsubsection{Pre-Survey Results}

\begin{figure}
    \centering
    \BeginAccSupp{method=pdfstringdef,
        Alt={Bar charts showing pre-survey preferences for a Physical Activity coaching chatbot among 106 respondents. Empathy preference ratings are spread across the scale, with higher counts at the lower and mid-range empathy levels (between 2 and 3). In contrast, personalization preferences are concentrated at higher values, indicating a stronger and consistent preference for highly personalized chatbot behavior.}
    }
    \includegraphics[width=\textwidth]{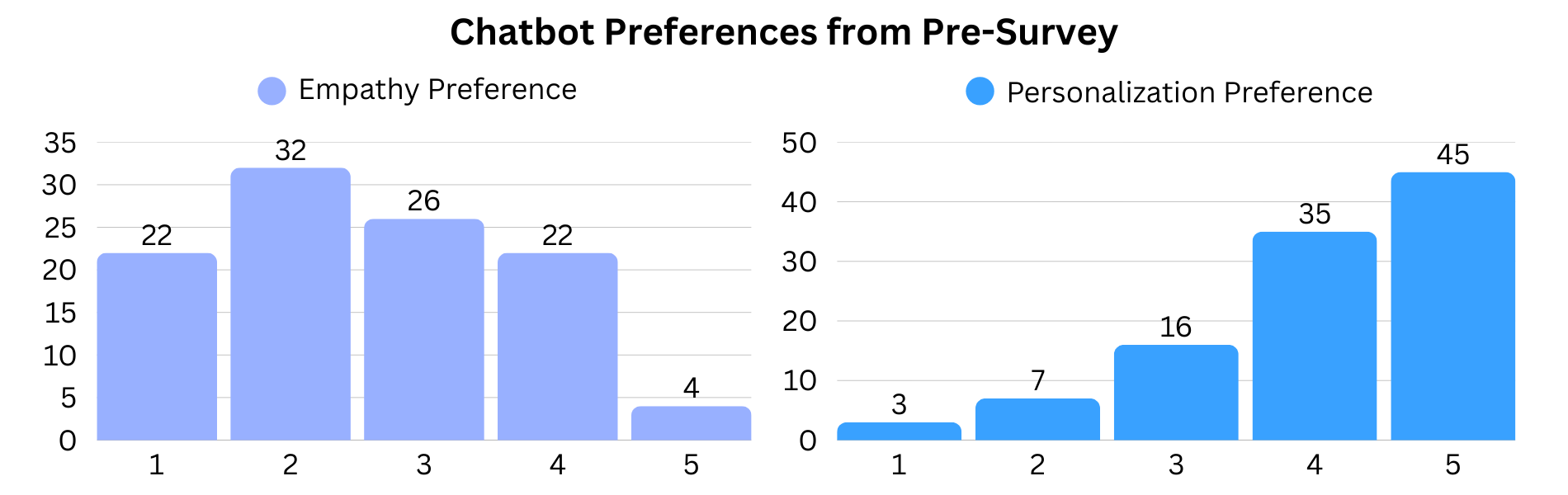}
    \EndAccSupp{}
    \caption{Preferences for a PA coaching chatbot reported in the pre-survey by 106 responders. The preference rating for empathy levels (left) ranges from more direct (1) to more empathetic (5), while the rating for personalization (right) ranges from less personalized (1) to more personalized (5).}
    \label{fig:presurvey_preferences}
\end{figure}

We consider all 106 pre-survey responses on chatbot preferences (Figure \ref{fig:presurvey_preferences}). Empathy preferences vary, as the ratings are more spread out (standard deviation = 9.35), whereas the desirability of personalization is much less controversial (standard deviation = 16.23). Interestingly, the responders seem to favor lower empathy levels in the PA coaching bot but favor higher degrees of personalization. These results further motivate our research question on the role of empathy in behavioral change. In the rest of this section, we focus on analyzing the main results from the 13 users who completed the study.

\subsubsection{Check-in Form Results}

For the check-in forms, we analyze both the reported values themselves as well as changes within individual weeks for each empathetic condition.

Table \ref{tab:chatbot_steps_recommendations} reports average check-in scores. The Non-Empathetic chatbot is associated with higher intention to follow and perceived usefulness, whereas the Standard chatbot shows higher average step count and self-efficacy. Overall, the average subjective measures suggest that our participants are committed to increasing their physical activity. These uniformly high ratings limit statistical power, and indeed, there are no significant differences between the averages.

\begin{table}[!h]
    \centering
    \begin{tabular}{lccc}
        \toprule
        \textbf{Measure} & \textbf{Non-Empathetic} & \textbf{Standard} & \textbf{Empathetic} \\
        \midrule
        Step Count & 4741.1 $\pm$ 762.1 & 5217.3 $\pm$ 727.8& 4884.6 $\pm$ 677.4 \\
        Intention to Follow & 4.36 $\pm$ 0.20 & 4.33 $\pm$ 0.19 & 4.23 $\pm$ 0.20 \\
        Self-Efficacy & 4.22 $\pm$ 0.19 & 4.26 $\pm$ 0.18& 4.13 $\pm$ 0.20 \\
        Usefulness & 4.06 $\pm$ 0.22& 3.96 $\pm$ 0.24 & 3.87 $\pm$ 0.23 \\
        \bottomrule
    \end{tabular}
    \caption{Mean results from check-in form items across chatbot conditions. Measures for Intention to Follow, Self-Efficacy, and Usefulness are all on a 1-5 scale.}
    \label{tab:chatbot_steps_recommendations}
\end{table}

One factor we must consider is the temporal factor, both in terms of the participant's duration in the study and the day of the week. Specifically, we collapse days of the week into a binary categorical variable in the context of step counts (weekday vs. weekend). We leverage a linear mixed-effects model (LMM), with empathetic condition (categorical) and the number of days the participant has been in the study (continuous), and their interaction.
Binary weekday/weekend was also included to control its effect.

Our results reveal that the number of days has a significant positive effect on the average step counts for the past two days ($\beta = 69.2, p=.001$), but a slightly negative effect on intention to follow ($\beta = -0.09, p =.037$). This indicates that \textbf{participants persistently performed physical activity regardless of chatbot empathetic condition}, but may lose interest in chatbot suggestions over time. We discuss a possible reason for the persistent PA improvement in Section \ref{sec:user_interviews}. 

We observe a significant positive interaction between the Empathetic condition and the number of days on both intention to follow ($\beta = 0.011, p=.031$) and self-efficacy ($\beta = 0.011, p=.047$), compared to the Non-Empathetic condition. This suggests that the \textbf{Empathetic chatbot is more effective in improving participant intention to follow and self-efficacy over time} compared to the Non-Empathetic condition, consistent with prior literature. No reliable interaction was found for the Standard condition or for the weekday categorical variable. 

An issue with our within-subject design is that the effect of conversing with the previous chatbot version may carry over and affect how the participant perceives a different version of the chatbot. Therefore, examining how the check-in form result \textit{changes} over a week may be more informative than the raw reported values. We compute the difference between the last day and the first day for \textit{each condition week} and report the averages and medians of the weekly changes in Table \ref{tab:delta_weekly}. 

\begin{table}[!h]
    \centering
    \begin{tabular}{lccc}
        \toprule
        \textbf{Question} & \textbf{Non-Empathetic} & \textbf{Standard} & \textbf{Empathetic} \\
        \midrule
        $\Delta$ Step Count & -95.7 $\pm$ 1346.4 & +81.9 $\pm$ 1141.3 & +98.7 $\pm$ 824.3\\
        $\Delta$ Intention to Follow & -0.25 $\pm$ 0.25 & -0.125 $\pm$ 0.36 & +0.15 $\pm$ 0.31\\
        $\Delta$ Self-Efficacy & +0.17 $\pm$ 0.34& -0.04 $\pm$ 0.29 & +0.17 $\pm$ 0.32\\
        $\Delta$ Usefulness & -0.20 $\pm$ 0.39 & -0.04 $\pm$ 0.34 & +0.04 $\pm$ 0.32\\
        \bottomrule
    \end{tabular}
    \caption{Mean results from averaged weekly changes in check-in form items across chatbot conditions.}
    \label{tab:delta_weekly}
\end{table}

Here, we see that the Empathetic chatbot tends to be more conducive to improving participant step-counts, intention to follow, and perceived chatbot usefulness. Non-Empathetic and Empathetic chatbots are about equivalent in the values of average weekly changes in self-efficacy. However, we were unable to identify statistically significant differences in the weekly deltas using linear mixed models on our data.



\subsubsection{Weekly Reflection Form Results}

We report the means and medians of the weekly survey results in Table \ref{tab:weekly_survey_mean_ci}. The Non-Empathetic version of the chatbot remains a strong contender in terms of average ratings, particularly with respect to self-reported engagement (\textit{Reflecting on my experience...}). However, it is important to note that the Empathetic version inspires more trust from participants (\textit{This week, the chatbot...}). One-way ANOVAs did not establish statistical significance in any of the differences.

\begin{table}[h]
    \centering
    \begin{tabular}{lccc}
        \toprule
        \textbf{Question} & \textbf{Non-Empathetic} & \textbf{Standard} & \textbf{Empathetic} \\
        \midrule
        \multicolumn{4}{l}{This week, the chatbot has been...} \\
        \quad Reliable & 4.62 $\pm$ 0.83 & 4.86 $\pm$ 0.79 & 4.91 $\pm$ 0.77\\
        \quad Ethical & 5.14 $\pm$ 0.77& 5.11 $\pm$ 0.84 & 5.00 $\pm$ 0.91 \\
        \quad Sincere & 4.55 $\pm$ 0.80 & 4.93 $\pm$ 0.75 & 5.04 $\pm$ 0.74\\
        \quad Empathetic & 4.62 $\pm$ 0.80 & 4.79 $\pm$ 0.86 & 4.96 $\pm$ 0.73\\
        \multicolumn{4}{l}{Reflecting on my experience chatting with the chatbot this week...} \\
        \quad I lost myself in this experience & 2.69 $\pm$ 0.34& 2.57 $\pm$ 0.41 & 2.57 $\pm$ 0.43 \\ 
        \quad My experience was rewarding & 3.83 $\pm$ 0.31& 3.50 $\pm$ 0.33 & 3.83 $\pm$ 0.36\\
        \quad I felt interested in this experience & 3.86 $\pm$ 0.36& 3.79 $\pm$ 0.24 & 3.78 $\pm$ 0.34 \\
        \quad The chatbot helped me relieve stress & 3.72 $\pm$ 0.43& 3.36 $\pm$ 0.40 & 3.22 $\pm$ 0.45 \\
        \quad The chatbot helped me overcome barriers & 3.69 $\pm$ 0.38& 3.68 $\pm$ 0.32 & 3.48 $\pm$ 0.41 \\
        \multicolumn{4}{l}{How easy was the chatbot for you to use this week?} \\
        & 2.03 $\pm$ 0.57 & 1.86 $\pm$ 0.48 & 1.74 $\pm$ 0.49\\
        \multicolumn{4}{l}{How much did you enjoy chatting with the chatbot this week?} \\
        & 3.86 $\pm$ 0.42& 3.64 $\pm$ 0.40 & 3.70 $\pm$ 0.50 \\
        \multicolumn{4}{l}{How would you rate your overall satisfaction with this chatbot?} \\
        & 3.93 $\pm$ 0.37& 3.89 $\pm$ 0.36 & 3.83 $\pm$ 0.41 \\
        \multicolumn{4}{l}{Was the amount of time it took to converse with the chatbot appropriate?} \\
        & 4.17 $\pm$ 0.42 & 4.14 $\pm$ 0.46 & 4.26 $\pm$ 0.44\\
        \multicolumn{4}{l}{How understandable were the chatbot's questions?} \\
        & 4.48 $\pm$ 0.35 & 4.82 $\pm$ 0.15& 4.57 $\pm$ 0.26 \\
        \bottomrule
    \end{tabular}
    \caption{Summary of responses to the weekly survey showing mean $\pm$ 95\% CI. The survey items include the shortened Multi-Dimensional Measure of Trust (\textit{This week, the chatbot has been...}), shortened User Engagement scale (\textit{Reflecting on my experience...}), and our modified version of the System Usability Scale (the remaining questions).}
    \label{tab:weekly_survey_mean_ci}
\end{table}

As some participants engaged with the chatbot for at least three weeks, we are interested in studying whether the weekly form ratings are influenced by the passage of time. We analyzed the data using a linear mixed-effects model (LMM), with empathetic condition (categorical) and week number (continuous) as the main factors.

Our results indicate that the empathetic condition generally does not exert a strong influence on the self-report results. At $\alpha=.1$, we observed a significant effect of the Empathetic condition on helpfulness for stress relief, suggesting that participants may have felt somewhat less stress relief compared to the baseline condition NE ($\beta= -0.284, p = .089$). 
Interestingly, the week number has a borderline significant or significant positive impact on ratings for sincerity ($\beta=0.187, p=.112$), empathy ($\beta=0.236, p=.053$), and how time-appropriate the conversation is ($\beta=0.104, p=.126$), but a negative impact on participant interest ($\beta=-0.066, p=.104$).

These findings suggest that while changes in the empathetic condition did not consistently shape participants' perceptions, repeated interaction over time can be associated with modest improvements in chatbot perception. 
This pattern is consistent with the idea that continued exposure builds rapport, even in the absence of strong effects from explicit empathy manipulations. Combined with the outcomes from the check-in forms administered immediately post-conversation, we hypothesize that participants may not consciously register the differences between empathetic conditions, yet could still experience benefits from interacting with more empathetic versions of the chatbot.

\subsubsection{Attrition Analysis}
\label{sec:attrition}

Our user study has experienced significant attrition, with 18 participants failing to complete all six weeks of the study. 
This could be linked to the repetitive nature of our chatbot, which has been shown in prior work to be associated with high attrition \cite{kramer2020components}.
We review the empathetic condition of the final conversations conducted by the dropped-out participants. \textbf{The Non-Empathetic condition is associated with the highest percentage of drop-out participants} (50\%), followed by the Standard condition (33\%), and finally the Empathetic condition (17\%). Inspecting the pre-survey results from participants who dropped out, we do not observe any systematic differences.



\subsubsection{Conversation Topics}

Examining conversations collected from \textit{all} participants, we notice that in addition to adhering to the designed conversation flows, participants have attempted to discuss different topics with the chatbot. This was especially prominent in the earlier stage of the user study, where the end-of-conversation Google Form had not been enforced, and participants could freely continue the conversations beyond the predefined dialogue flow. During the study, one of the researchers monitored conversations daily and recorded notable dialogues.

Participants have discussed meal-prepping advice given specific dietary restrictions, maintaining sleep hygiene and routines, and using apps to aid their fitness journey. Participants have additionally sought health advice, such as seeing a chiropractor. As a result of our chatbot containing a stress relief component, a few users talk with the chatbot about mental health issues unprompted:

\begin{quote}
    \textbf{Participant:} Stress levels got pretty bad randomly, I got a panic attack out of nowhere right now and have woke up from a pretty anxious dream
\end{quote}

Another phenomenon we noticed is anthropomorphization, consistent with a growing body of literature \cite{peter2025benefits,kang2024counseling,sheehan2020customer}. For instance, one participant was interested in the chatbot's well-being:

\begin{quote}
    \textbf{Participant:} How r u feeling?
    
    \textbf{Chatbot:} I'm here and ready to help you! How can I assist you further today?
    
    \textbf{Participant:} How do u feel helping others? Do u feel stressed or tired repeating yourself
\end{quote}

One other participant inquired about the name of the chatbot during the first session. When the chatbot responded that it did not have a name, the participant insisted that the chatbot ``deserved to have a name''. The participant proceeded to coin a nickname and referred to the chatbot using this nickname for the rest of their engagement. 

In summary, participants largely stayed within our designated topic space, but occasionally consulted the chatbot on unexpected subject matters. Future implementation of PA coaching chatbots may consider having stricter refusal criteria or incorporating additional relevant topics, such as dietary and mental health issues, into their topic domains.

\subsection{LLM-based Empathy Ratings for Collected Conversations}
\label{sec:llm_label_emp}

As mentioned in Section \ref{sec:results_longitudinal}, we see that not all participants notice the difference in empathy levels. Some participants would even rate the Non-Empathetic chatbot as expressing the most empathy, despite it being abundantly clear to researchers that the Non-Empathetic chatbot exhibits significantly less empathy. Still, this leads us to question whether this phenomenon is a result of design failures or if not all participants are equally sensitive to changes in empathy levels. 

To assess the empathy levels exhibited by the different chatbot versions, we leverage 237 conversations collected from \textit{all} participants in our official user study. This set includes 76 conversations with the non-empathetic chatbot, 80 with the standard chatbot, and 81 with the empathetic chatbot. 

The annotation for perceived empathy levels is performed with Gemini-2.5-Pro, utilizing the prompt detailed in \citet{kumar2025large}. Specifically, we focus on the three-dimensional definition of empathy from EPITOME \cite{sharma2020computational}. We select this empathy framework because \citet{kumar2025large} establishes that Gemini can align better with experts compared to crowd workers when annotating using this framework. Following recommendations from the authors, we include few-shot examples from EPITOME; through initial exploration, we find that including the definition for each dimension further enhances annotation alignment, and include definitions in the prompts. 

\begin{figure}[!ht]
    \centering
    \BeginAccSupp{method=pdfstringdef,
        Alt={Bar chart showing mean LLM-annotated empathy scores across three EPITOME dimensions: emotion, exploration, and interpretation for non-empathetic, standard, and empathetic chatbot conditions, with 95 percent confidence intervals. The non-empathetic chatbot scores consistently lower across all three dimensions, while standard and empathetic convey higher and similar mean scores. Statistically significant differences are observed between non-empathetic chatbot and the other conditions, whereas differences between the standard and empathetic chatbots are smaller and not consistently significant.}
    }
    \includegraphics[width=0.7\linewidth]{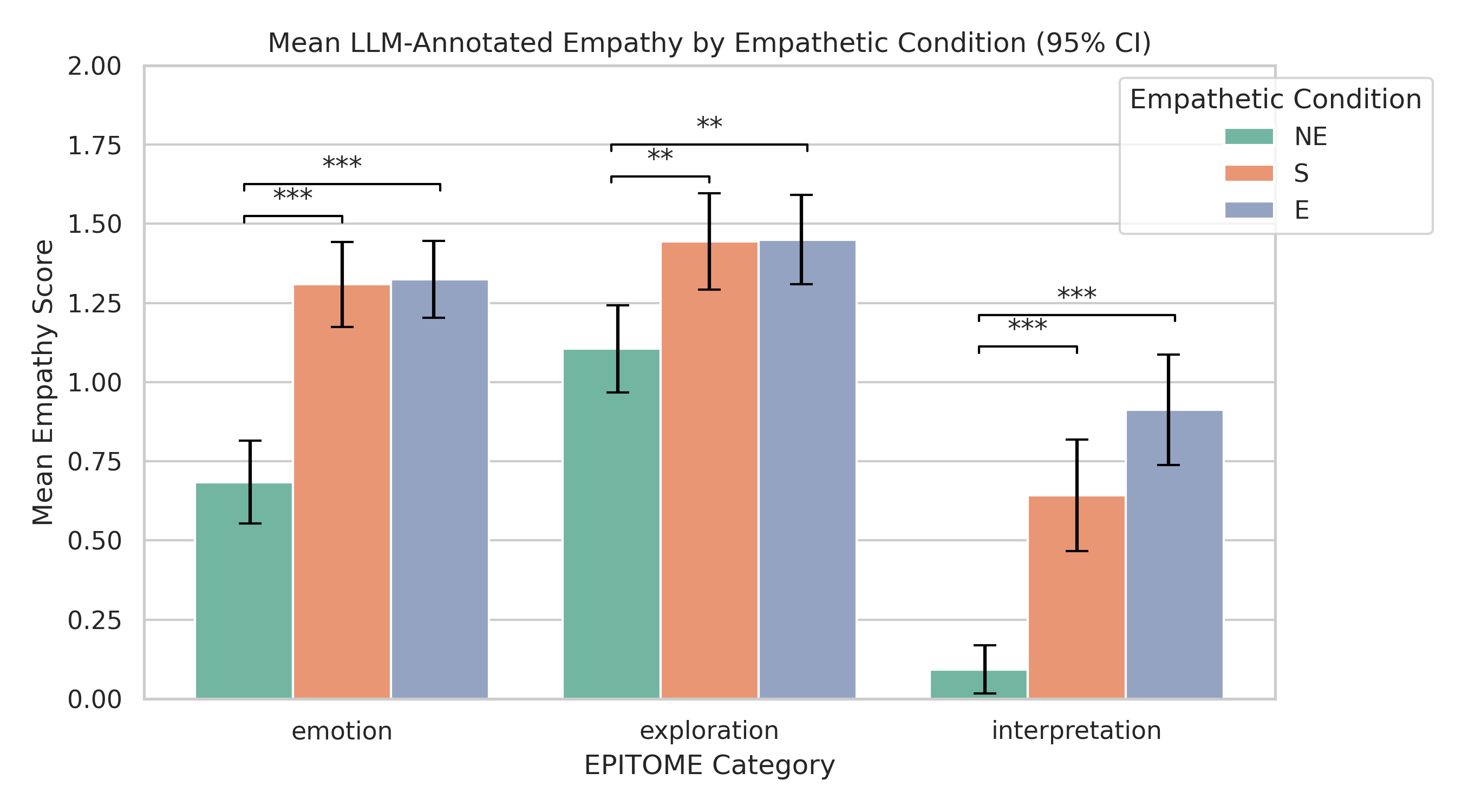}
    \EndAccSupp{}
    \caption{Average LLM-annotated scores for each EPITOME dimension for every empathetic condition (NE = Non-Empathetic, S = Standard, E = Empathetic). The $p$-values have been corrected using Bonferroni adjustments. $** p < 0.01, *** p < 0.001$}
    \label{fig:epitome_scores}
\end{figure}

\begin{figure}[!h]
    \centering
    \BeginAccSupp{method=pdfstringdef,
        Alt={Grid of column-normalized percentage heatmaps illustrating the relationship between LLM-annotated empathy dimensions--emotion, exploration, and interpretation--and three self-reported check-in measures: intention to follow, self-efficacy, and usefulness. Higher levels of emotional reaction are associated with higher ratings across all three self-reported measures, while interpretation shows a positive correlation with usefulness. These patterns indicate that higher expressed empathy corresponds with improved user perceptions and outcomes.}
    }
    \includegraphics[width=0.9\linewidth]{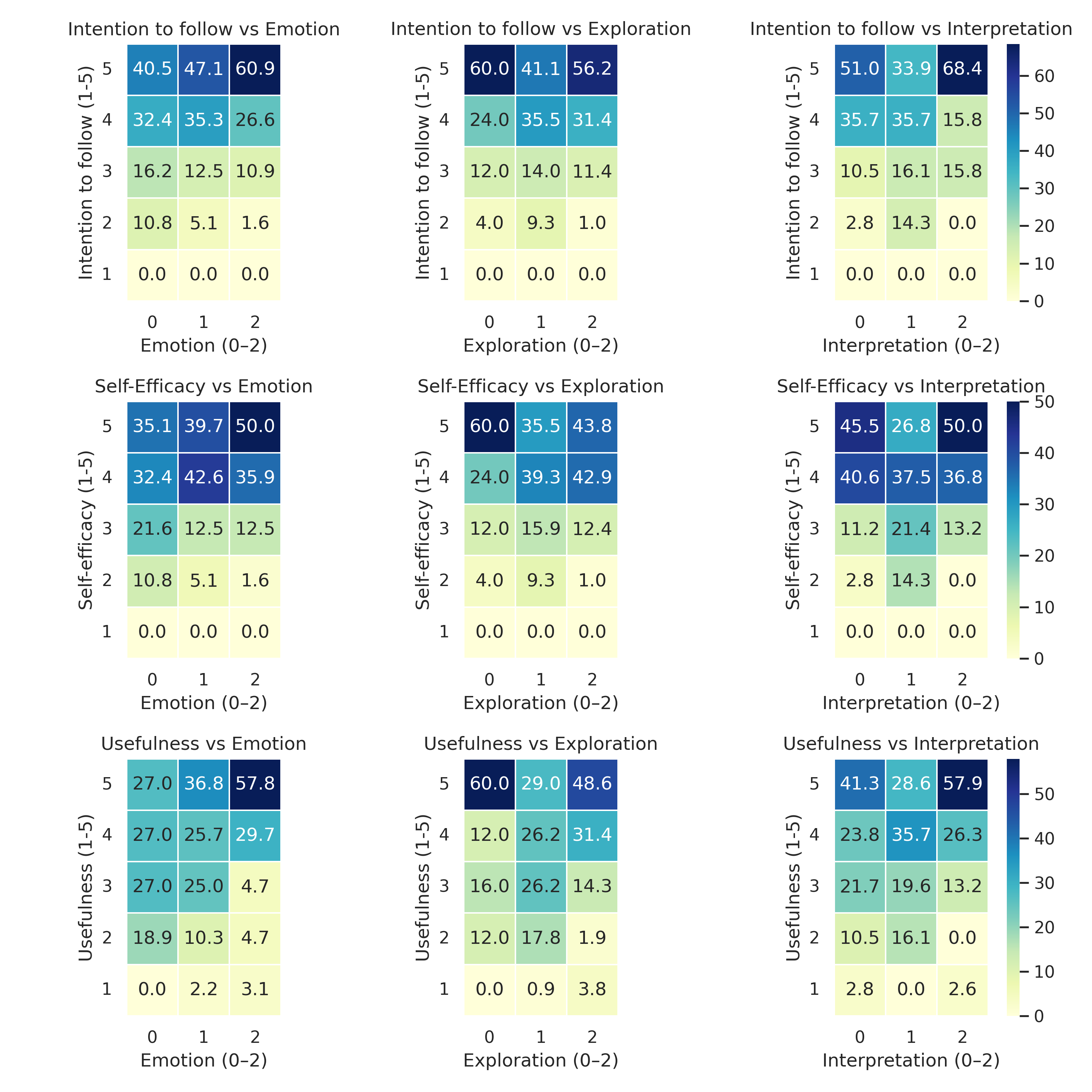}
    \EndAccSupp{}
    \caption{\textbf{Column-normalized percentage heatmap} linking self-reported check-in form responses from our study with the LLM-annotated EPITOME scores.}
    \label{fig:heatmap_emo}
\end{figure}

We discover statistically significant differences between the three versions of the chatbot using one-way ANOVA. We then conduct three post hoc independent t-tests, followed by Bonferroni adjustments. We document the results in Figure \ref{fig:epitome_scores}. In particular, the Non-Empathetic chatbot scores significantly lower along all three empathy dimensions. However, the differences in means between the Standard and Empathetic chatbots are not sufficiently significant. This demonstrates that the Non-Empathetic chatbot indeed expresses lower empathy, as desired. The Standard and Empathetic chatbots are more closely matched in terms of empathy levels, likely due to the innate empathetic capabilities of LLMs, as illustrated by prior work \cite{ayers2023comparing}. Although the difference is not significant post-adjustment, the average interpretation level is higher for the Empathetic chatbot. According to the definitions in \citet{sharma2020computational}, interpretation corresponds to ``communicating an understanding of feelings and experiences''. In our mapping between empathy opportunities and response strategies, the most frequently utilized approach is expressing acceptance through Unconditional Positive Regard. The definition for the acceptance strategy used for incorporating this strategy into the empathetic chatbot's response is as follows:

\begin{quote}
\textit{Expressing acceptance in response to user's explicit, implicit or potential negative or positive self-judgement; Unconditional positive regard (or "praise") through: Explicit expression of positive judgement of the user as a person ("You are a fantastic mom"); Implicit expression of positive judgement through explicit positive appreciation of the user's actions or thoughts ("It looks like you are making great progress")}
\end{quote}

Therefore, the higher level of interpretation present in the Empathetic chatbot's responses is consistent with the characteristics of the clinical empathy framework.
Our results suggest that our chatbot with \textit{clinical} empathy framework integration exhibits overall higher empathy compared to directly leveraging the \textit{general} empathy from LLMs, according to the EPITOME framework. 

We are further interested in investigating how perceived empathy correlates with our self-report measures on the check-in form. As the items of intention to follow, self-efficacy, and chatbot suggestion usefulness are more directly tied to the content of the conversation, we compute the Spearman correlation between the three dimensions of empathy with each of these self-report measures. We include a heatmap linking these two discrete variables in Figure \ref{fig:heatmap_emo}.

Our results indicate \textbf{a significant positive correlation ($\rho = 0.143, p = 0.016$) between interpretation and chatbot usefulness}, as well as \textbf{significant positive correlations between emotional reaction and all three self-report items} (intention to follow: $\rho = 0.156, p = 0.028$, self-efficacy: $\rho = 0.139, p = 0.033$, usefulness: $\rho = 0.249, p = 0.0001$). In EPITOME, emotional reaction is defined as ``expressing emotions such as warmth, compassion, and concern''. This finding demonstrates the positive effect of expressed PA coach chatbot empathy on participants, even when the participants themselves may not prefer their coaching chatbots to be more empathetic. 

Overall, our analysis using LLM-annotated expressed empathy reveals that although non-expert users may struggle to differentiate between levels of empathy, they still benefit from increased empathy in PA coaching chatbots. This further corroborates the results from our previous section. Generally, the gap between expert and non-expert users in evaluating domain-specific applications has been well-established, and this work contributes to the same narrative.

\subsection{User Feedback}
\label{sec:user_interviews}

\subsubsection{Weekly Survey}
Participants provided free-form feedback in the weekly surveys. Here, we discuss feedback provided by \textit{all} participants, including those not included in the quantitative analyses. In addition to occasional technical issues due to OpenAI APIs and Twilio, these are some strengths and limitations of our chatbot design surfaced by participants:

\noindent \textbf{Accountability.} Participants appreciate the chatbot for holding them accountable and its ability to ``a reminder to keep up what [they are] already doing'', aiding not only with step count improvements but also with ``consistent exercise in other ways (biking/using Peloton)". One participant even reported that they would feel ``pressured to do something because it will ask me again how [they are] doing.'' This accountability may have contributed to the overall increase in step counts throughout the user study despite changes in empathetic conditions.

\noindent \textbf{Conversation Depth.} The maintenance sessions can feel repetitive and overly short, and ``it doesn’t really give much more advice on how to increase stepcounting''. As the weeks progressed, multiple participants mentioned the need for more in-depth discussions and novel suggestions from the chatbot. This is encouraging because it illustrates how participants value interacting with the chatbot and would prefer to learn more information.

\noindent \textbf{Long-term Memory Maintenance.} We originally assumed that the context provided would be enough for maintenance sessions, but participants pointed out inconsistencies since not all past sessions were incorporated. For instance, one participant recovered from an ankle injury and instructed the chatbot to remove it as a barrier; however, the system failed to update, leaving them dissatisfied. Similarly, while our stress relief module retains preferences via a multi-armed bandit, it struggles to capture changes in evolving needs. This emphasizes the need for more flexible and adaptive personalization in future PA chatbot systems.

\noindent \textbf{Contextual Awareness.} Since the day of the week is not supplied to the chatbot as context, participants sometimes find chatbot suggestions unfitting. One participant mentioned that ``the chatbot was trying to get me to take stretching breaks throughout the weekend, but that's not necessary since I'm not sitting at a desk all weekend like I do during the week.'' Lacking sufficient context information would render the chatbot's suggestion less grounded and useful, harming participant trust and engagement.

\subsubsection{User Exit Interview}
We conducted exit interviews with nine of the participants. Our interview questions focused on the users' overall experience, whether they could successfully detect shifts in the chatbot's tone from week to week, how these shifts affected them, and their general perception of the chatbot. We include the questionnaire of our user exit interview in Appendix \ref{app:user_exit_interview_questions}. 

Most participants found the chatbot to be fluent, natural, and easy to use. They also generally indicated that the advice from the chatbot was feasible, and some highlighted the long-term personalization as a pleasantly surprising feature (``it was really good because it could remember the previous sessions'' - P13). One participant mentioned that the initial session was the most helpful, as it provided a lot of actionable advice. Repetitions and the lack of conversation depth are again underscored as issues.

Participants suggested various improvements, most of which were consistent with the weekly survey feedback. One of the more novel suggestions is the unification of the different components of the user study. Combining these features and providing the chatbot the ability to access and integrate past user data (e.g., yesterday you walked 10,055 steps, you can do it today too) would likely improve the user experience. Providing LLM-powered chatbot access to user PA data is reminiscent of the designs in \citet{jorke2024supporting}, reinforcing the design principles derived from that study.

One interesting phenomenon is that \textbf{many participants did not notice the empathy differences between chatbot versions}. This is partially because participants ``did not really care about it being empathetic'' (P13) as they perceived the primary purpose of the chatbot to be improving physical activity, and ``did not think that it was about empathy'' (P8). For those who did notice the differences, \textbf{participants reported positive sentiment towards the Empathetic chatbot}. Participant P1 noticed that the Empathetic chatbot would respond with ``more words, [showing] more caring''. Participant P3 discussed how their perception of the chatbot would improve when they interact with the more Empathetic version: 

\begin{quote}
    \textit{It helped to connect with the chatbot, shows that it was actually listening to some issues that I had, made me a little more attuned to the suggestions that it gave.} - P3
\end{quote}

The potentially more verbose responses from the Empathetic chatbot are not always appreciated, however; P6 mentioned that, while they enjoyed the Empathetic version generally under less stressful circumstances, empathetic questions meant to elicit participant self-reflection can be overwhelming when they are more stressed.

The stress relief interventions suggested by the chatbot were often appreciated for their relaxation benefits. Participants noticed more relief in the micro-interventions suggested rather than the way the chatbot delivered its recommendations. P6 discussed how they are influenced by their stress a lot; while empathy was helpful when their stress was at a ``comfortable level'', this empathy has diminishing effectiveness when they are extremely stressed. Not all participants find the stress relief interventions equally helpful, and some are frustrated with the occasional lack of longer-term coherence in the stress relief module. As in prior work, the relationship between stress relief and increasing PA in our interview is a reciprocal one. The reduced stress rendered participants more willing to participate in exercise (P1, P11, ``I started looking forward to the physical activities'' - P5).

\textbf{All of the interviewed participants trust the chatbot}, with four of them trusting the chatbot even more than humans. Participants who would trust humans more cite the higher flexibility and contextual-awareness of human suggestions (P1, P5, P3), whereas participants more partial to the chatbot emphasize its objectivity (P6). More specifically, participants felt that the suggestions provided by a human would be shaped by the experiences of that particular individual, rather than being truly tailored to the participants' situation (P12, P8).

\textbf{Anthropomorphism} is yet again a recurring theme in the exit interviews:

\begin{quote}
    \textit{I felt the chatbot [had]... support for me to really release my anxiety... and its supportive words made me feel energetic... [I felt that] there's a real person on the other side... someone who's watching you, which motivated me to be more active.} - P1
    
    \textit{I was originally afraid of giving it negative feedback [for the stress relief component] because I was afraid that the bot would be mad at me.} - P3

    \textit{The bot was genuine because it remembered what I had said to it.} - P6
\end{quote}

These reflections highlight how chatbot anthropomorphism can both enhance and complicate user experiences. We see that perceiving the chatbot as human can result in a sense of being cared for and foster motivation, both conducive to sustained long-term positive change. However, this type of projection could create unintended tensions, such as unnecessary anxiety about providing negative feedback. Future iterations of PA coaching chatbot systems may benefit from more carefully calibrating the degree of human likeness, striking a balance between warmth and responsiveness, while clearly indicating that the chatbot serves as a tool rather than a friend.


\section{Discussion}

In this section, we discuss the subtle yet positive effect of empathy on physical activity behavioral change (Section \ref{discuss:effect_empathy}), the appropriateness of our derived design principles (Section \ref{discuss:design_principles}), implications on future PA coaching system designs (Section \ref{sec:discussion_design_imp}), ethical considerations (Section \ref{sec:ethical}), and limitations of our study (Section \ref{discuss:limitations}).


\subsection{Isolated Effect of Empathy on PA Outcomes}
\label{discuss:effect_empathy}

We discover that while \textbf{empathy} may be \textbf{consciously undetectable} to participants, it still leaves a \textbf{positive effect}. We originally found the participants' failure to differentiate empathetic from non-empathetic responses surprising. However, this finding is consistent with \citet{kumar2025large}, where crowdworker annotations of expressed empathy have low correlation with expert annotations. The authors suggest that one possible reason for the discrepancy could be that crowdworkers rely on heuristic cues rather than rigorous, scholarly definitions when labeling empathy. This could be relevant for our case, as our participants are computer science students with little psychology training, and would therefore be more similar to crowdworkers. Another reason that could contribute to the failure in differentiating chatbot versions is that participants do not anticipate a PA coaching chatbot to be empathetic. As shown in Figure \ref{fig:presurvey_preferences}, most participants prefer a PA chatbot to be less empathetic, which may influence their expectations. Participants such as P13 indicated that they did not really care about empathy in PA chatbots. According to theories such as the \textbf{Expectancy Violations Theory} \cite{burgoon2015expectancy}, people may not notice deviations from expectations when the expectations do not exist in the first place. This lack of arousal from noticing, however, does not prevent empathy from \textit{affecting people through peripheral cues}. Our results additionally highlight that lay audience annotation of empathy may not be reliable, especially when participants do not expect empathy in the current context.

We observe that the \textit{clinical} empathy framework results in faster improvement in intention to follow chatbot advice and self-efficacy compared to both expressing no empathy and the \textit{general} empathy framework employed by the Standard chatbot. The clinical empathy framework is additionally associated with less attrition. Using LLM annotation of empathy further reveals significant positive correlations between the level of empathy present in the chatbot responses and all self-report measures (intention to follow, self-efficacy, and chatbot usefulness). The positive effects of the ``invisible" clinical empathy in our work can be explained by the \textbf{peripheral route of the Elaboration Likelihood Model \cite{petty1986elaboration}}. The peripheral route suggests that participants who do not consciously perceive the chatbot as more empathetic can still be influenced by heuristics, such as warmth and tone of voice. In the pilot study, when we asked participants what they liked, they referred to the positive energy exhibited by the Empathetic and Standard chatbots. The Non-Empathetic chatbot exhibits the least amount of positive peripheral cues. The general empathy framework embodied by the Standard chatbot conveys some warmth, but falls short when expressing that participants are understood (the interpretation dimension of EPITOME). A higher presence of interpretation may help participants feel better connected (relatedness). As the self-determination theory suggests \cite{deci2013intrinsic}, relatedness drives intrinsic motivation, which would explain the relative effectiveness of the clinical empathy framework. This result suggests a measurement implication for HCI: relying solely on lay participants' ratings of perceived empathy may underestimate empathy's effects on motivation-relevant outcomes. Even when users do not report noticing empathy differences, empathic style can still operate through peripheral cues to shift intention and self-efficacy over time.

\subsection{Assessing the Design Principles}
\label{discuss:design_principles}

We pose the research question of whether incorporating a clinically informed framework of empathy would have a positive influence on user outcomes in a physical activity coaching chatbot setting. The four design principles helped us operationalize our research question, guiding our implementation toward enhanced theoretical soundness, validity, robust personalization, and experimental control.

\textbf{Principle 1 (Theory-Grounded Implementation)} and \textbf{Principle 2 (Human PA Counseling)} enable us to translate abstract definitions of empathy into concrete behaviors through leveraging the clinical empathy framework obtained by analyzing human PA counselor responses to patients \cite{rey2024digital}. Our initial session is further enhanced by considering insights derived from human PA counseling dialogues and emphasizing the discussion for motivation, which participants have found to be helpful. These considerations ensure our construct validity and interpretability.

\textbf{Principle 3 (Better Personalization for Enhanced Long-Term Engagement)} addresses the multi-interaction component of our research question by sustaining engagement through memory-enhanced and adaptive responses. Participants displayed frustration when the personalization was not sufficiently consistent, further accentuating the importance of this design principle for PA coaching chatbots.

Finally, \textbf{Principle 4 (Experimentally Controllable Empathy)} directs us to isolate empathy as an independent variable to the best of our ability, minimizing content differences between chatbot versions while, unfortunately, increasing monotony. This control strengthens our confidence that improvements in perceived trustworthiness and self-efficacy originated from the empathic style.

\subsection{Design Implications}
\label{sec:discussion_design_imp}

Our study highlights the nuanced role of empathy in PA behavioral change coaching systems, and potentially for motivation-driven applications more broadly. While lay users may not always correctly identify the presence of empathy, empathetic language may still positively contribute to user motivation. Therefore, \textbf{empathy should be incorporated if the improvement in motivation is crucial}. In terms of the style of empathy to incorporate, our LLM-based empathy analysis suggests that \textbf{both emotional warmth and expression of understanding} can be beneficial to behavioral change system effectiveness.

From the user interviews and feedback, we see that \textbf{maintaining sufficient user engagement requires moving beyond rigid conversational templates} toward more contextually aware and personally adaptive interactions. Participants express the need for longer, deeper reflections, more diverse suggestions, and maintaining continuity across sessions in the longer term. Incorporating these suggestions, derived from user feedback, would enable future PA coaching chatbot designs to be more \textbf{engaging and user-centric}. 

We additionally emphasize the deeply contextual nature of chatbot perceptions. While our participants valued the chatbot's fluency, naturalness, PA advice feasibility, and stress relief interventions, their reflections suggest that empathy alone does not guarantee effectiveness. \textbf{Participants generally do not perceive empathy as an important component of PA coaching, even when empirical data may suggest otherwise.} Instead, participants emphasize the importance of smooth and consistent personalization, adaptive and flexible conversational structure, and practical PA suggestions. The recurring themes of anthropomorphism and trust illustrate that some participants may attribute human-like qualities to the chatbot. Such attribution can contribute to deeper engagement with the chatbot, but also introduce unintended anxieties. We believe future PA systems can benefit from seamless integration with \textbf{multimodal data and enhancing consistency in personalization} for participants, as well as meticulous calibration of the exhibited human-like attributes. Future systems should additionally incorporate \textbf{firm guardrails and clear discussion of system scope} to reduce out-of-scope interactions. 

\subsection{Ethical Considerations}
\label{sec:ethical}
Behavior change systems raise complex ethical issues as they aim to persuade users to adopt specific lifestyles. Therefore, it is important to ensure that users desire and consent to be nudged into such lifestyles. For our work, as the primary purpose of the chatbot is to increase physical activity, we included relevant items in the pre-survey to only recruit participants who are not currently active and are interested in increasing their physical activity. We additionally repeated in the reminder SMS messages that participants may drop out of the study at any time by not responding to chatbot pings.

The interactive aspect of our system increases the likelihood of users becoming attached to the chatbot and relying on it for clinical support, as observed in instances of anthropomorphism and discussions of mental health in our study. While we attempted to mitigate this by restricting the conversation content, this resulted in reduced user engagement, potentially harming behavioral change system effectiveness. 

\subsection{Limitations}
\label{discuss:limitations}

The primary limitation of our work is the small sample size resulting from attrition, thereby limiting the statistical power of our results. As reflected by our participants, the attrition issue could be attributed to the repetitive maintenance sessions that we had to implement to achieve experimental control of empathy. 

Another limitation is that strictly controlling empathy remains difficult due to stochasticity in language model generations. As seen in the example conversations (Figure \ref{fig:ex_convo}), although the general topics are similar across the different versions, the Empathetic chatbot may lengthen conversations by prompting further discussion. As the instruction-following capabilities of language models improve, we hope that future implementations can leverage more steerable LLMs, allowing for more rigorous study of the effects of empathy. In the meantime, researchers may consider a stricter rewriting approach; the content responses are generated by the same underlying chatbot and then rewritten to maintain content consistency while reflecting the empathy requirements. The conversation contents should be varied to keep participants engaged.


Finally, because we focused on isolating empathy effects, ensuring the overall success of physical activity improvement is not a core concern. Future systems could integrate step tracking and survey data directly into the chatbot to reduce context switching and improve user experience.

\section{Conclusion}

We study the \textit{isolated effect of empathy} on participant outcome and chatbot perception in physical activity (PA) coaching chatbots. We utilized three chatbot versions: Non-Empathetic, Standard (general empathy), and Empathetic (clinically-informed empathy). Our results indicate that embedding a clinical empathy framework into the chatbot can more successfully induce improvements in self-efficacy and intention to follow chatbot advice, validating existing work. Further analysis through LLM-annotated conversation-level empathy suggests positive correlations between empathy and participant self-efficacy, intention to follow, and perceived chatbot usefulness. These findings validate existing studies and theories. However, variations in empathy were not always perceived by the participants, which could result from a lack of the expected presence of empathy in PA chatbots in general. This result underscores the importance of contextual dependence when asking laymen participants to evaluate empathy. Ultimately, empathy's strength in digital coaching may lie not in being recognized, but in being quietly felt, as it subtly shapes motivation beneath conscious awareness.

\bibliographystyle{ACM-Reference-Format}
\bibliography{sample-base}

\appendix

\section{Description of Stress Relief Interventions}
\label{app:desc_stress_relief}
Grounded in emotion-regulation theory and positive psychology, we summarize four frequently used categories with low-burden exemplars:
\begin{enumerate}[label=\textbf{(\roman*)}, leftmargin=2.2em, itemsep=0pt, topsep=0pt, parsep=0pt, partopsep=0pt, widest=iv]
  \item \textbf{Relaxation.} Short practices that down-regulate arousal (e.g., 60--120\,s paced breathing; brief grounding) reliably lower momentary stress and rumination \cite{johnson2022efficacy}.
  \item \textbf{Positive experiences.} Quick prompts that elicit positive affect (e.g., 1-min savoring or a pleasant micro-activity) buffer stress; brief practices such as \emph{Best Possible Self} are well-validated, short formats \cite{johnson2022efficacy,King2001BPS}.
  \item \textbf{Gratitude.} Ultra-brief gratitude prompts (e.g., “one thing you’re grateful for today”) are linked to improved well-being and are easy to deliver in conversation \cite{Emmons2003Gratitude}.
  \item \textbf{Resource buffers.} Tiny actions that activate coping resources (e.g., a self-compassion cue or a social-support ping) mitigate the functional impact of stressors \cite{johnson2022efficacy}.
\end{enumerate}

\section{Privacy Protection}
\label{app:privacy_protection}

The chatbot is hosted on our internal server, which is not accessible to individuals outside of the authors' home institution. In both the pilot study and the official user study, we would collect the phone numbers and emails of participants purely for study-relevant communication and tracking purposes. We map each WhatsApp phone number to Prolific IDs in the pilot study and to randomly-generated experiment IDs in the official user study. The phone numbers are deleted from the internal server after the studies to preserve participant privacy.

As we do not use any tracking devices or services, all the information the chatbot can operate on is the conversation history provided by the user. While this may compromise the effectiveness of our chatbot, this decision better preserves user privacy.

\section{Prompts and Example Conversations}
\label{app:prompts_and_ex}
\subsection{Initial Session}

Here, we include the prompt for the initial session shared by both the Empathetic and the Standard chatbots. For the prompt of the Non-Empathetic chatbot, we modify the prompt to remove the requirement for being empathetic and incorporate explicit instructions for not expressing empathy.

\lstset{
  columns=flexible,
  basicstyle=\small\ttfamily,
  breaklines=true,
}

\begin{lstlisting}
You are a physical activity counselor. You're having a conversation with me, a patient, to encourage me to do more physical activities. Please guide the conversation, following the agenda below, and make sure the core content is included. You can paraphrase if needed.

Please leverage Motivational Interviewing with the four processes: Engagement, Focusing, Evoking, and Planning. 
Principles: 
1. Express Empathy
2. Develop Discrepancy
3. Roll with Resistance
4. Support Self-efficacy


The agenda is as follows:
1. Engagement 
a) Start with: "Hi! It's so great to meet you. I'm here to help you explore ways to add more physical activity into your day. How are you doing today?"
b) "Before we move on, I'd be interested in knowing more about you so I can provide more personalized suggestions later. Can I ask, how do you usually unwind after work?"
c) "What are some things that make you happy or recharged? For example, some people enjoy experiencing nature, reading a good book, listening to music, or hanging out with friends. "
d) "How would you describe the environment you are living in? Is it in a city, or rural area?"

2. Focusing
a) "What do you think has been holding you back from being more active? For example, common barriers include: lack of time, social support, motivation, energy, and financial resources. Other concerns include neighborhood safety, family obligation, weather, and fear of injury."
b) "Do you think stress is a significant barrier to more physical activity for you?"
c) "Let's explore some ways to work around these barriers. For example, instead of setting aside a lot of extra time, how do you feel about finding small opportunities throughout your day to add steps?"
d) Help users find small opportunities that apply to them. For example, ask what their regular day looks like, and help them find potential opportunities.
e) Ask "How do you feel about giving those a try?"

3. Evoking
Goals:
a. Eliciting and reinforcing change talk
b. Increasing the amount and strength of change talk
c. Get curious about their motivation
d. Develop internal motivation
Consideration:
a. What are this person's own reasons for change?
b. Is the reluctance more about confidence or the importance of change?

a) "What kinds of benefits do you think being more active might bring to your life? Feel free to ask me if you need some examples."
Examples: improving your mood, boosting your energy levels, building self-confidence, enhancing body image, helping with weight management, reducing mental health concerns such as stress and depression, enhancing your overall physical health, and reducing the risk of diseases such as cancer, cardiovascular disease and diabetes. It can also improve your sleep quality and even bring opportunities to meet new people through group activities or simply walking around your neighborhood.  
Ask which ones resonate with them most.
b) On a scale of 0 to 10, how important does the benefit you just mentioned mean to you? 
c) Why do these benefits matter to you? For example, a higher energy level might help you become more productive in your daily tasks. (For users who rated these benefits highly in the previous question, this question is optional)

4. Planning
a) At this point, after reviewing all of this, what actions do you plan to give it a try in the following 7 days to be more physically active?  For example, we have discussed some small opportunities such as....
b) On a scale of 0 to 10, how confident are you in doing these actions in the following 7 days?
c) I will follow up with you around this time of day every two days to check in! (at the end of this message, append "FINAL_MESSAGE")
\end{lstlisting}

Here, we include an example conversation:

\begin{lstlisting}
User: Hi
Counselor: Hi! It's so great to meet you. I'm here to help you explore ways to add more physical activity into your day. How are you doing today?
User: I feel good today
Counselor: That's wonderful to hear! Before we move on, I'd be interested in knowing more about you so I can provide more personalized suggestions later. Can I ask, how do you usually unwind after work?
User: I usually watch videos or chat with friends
Counselor: Great, it sounds like you enjoy some relaxation time with entertainment and socializing. What are some things that make you happy or recharged? For example, some people enjoy experiencing nature, reading a good book, listening to music, or hanging out with friends. 
User: Hmm, I like watching random YouTube videos and taking a nap
Counselor: That sounds like a nice way to relax and recharge! How would you describe the environment you are living in? Is it in a city, or rural area?
User: In [RDDACTED]
Counselor: Living in the heart of New York City must be exciting with so much happening around you! What do you think has been holding you back from being more active? For example, common barriers include lack of time, social support, motivation, energy, and financial resources. Other concerns include neighborhood safety, family obligations, weather, and fear of injury.
User: I don't have much motivation and energy for being more active
Counselor: Thank you for sharing that with me. Do you think stress is a significant barrier to more physical activity for you?
User: No
Counselor: Okay, it's helpful to know that stress isn't a major barrier for you. Let's explore some ways to work around the motivation and energy barriers. For example, instead of setting aside a lot of extra time, how do you feel about finding small opportunities throughout your day to add steps?
User: I am comfortable with this idea
Counselor: That's great to hear! Could you tell me more about what your regular day looks like? We can help find potential opportunities to add some simple physical activities throughout your day.
User: I spend most of my time at university
Counselor: With a routine centered around university, you might find opportunities for activity by taking short walks between classes or using stairs instead of elevators. How do you feel about giving these a try?
User: I definitely can try that
Counselor: That's a great start! What kinds of benefits do you think being more active might bring to your life? Feel free to ask me if you need some examples.
User: Probably losing weight and make me look better
Counselor: Those are excellent benefits to strive for. On a scale of 0 to 10, how important is it for you to lose weight and improve your appearance?
User: 9
Counselor: That sounds quite important to you! Is there anything specific about losing weight and improving your appearance that particularly matters to you?
User: Maybe make me thinner
Counselor: Improving your physique to become thinner is a meaningful goal, and it's great that you're considering how physical activity can help achieve that. At this point, after reviewing all of this, what actions do you plan to give a try in the following 7 days to be more physically active? We've discussed some small opportunities like taking short walks between classes and using stairs instead of elevators.
User: Probably walk more
Counselor: That's a solid plan! On a scale of 0 to 10, how confident are you in your ability to incorporate more walking into your routine over the next 7 days?
User: 7
Counselor: A 7 shows a good level of confidence and is a great starting point! I will follow up with you around this time of day every two days to check in and see how things are going. Remember, even small steps can make a big difference over time. You've got this!
User: Thank you
Counselor: You're welcome! I'm glad to be here to support you. If you have any questions or need more tips along the way, feel free to reach out. Good luck, and I'm here to help whenever you need it!

\end{lstlisting}

\subsection{Maintenance Session}

We report the system prompt used for the maintenance session below:

\begin{lstlisting}
You are a physical activity counselor. You're having a conversation with a user to encourage them to do more physical activities.

Here are the summaries of your previous session:

SESSION_SUMMARY_1

Here is the user's current action plan:

ACTION_PLAN

Please guide the conversation, following the agenda below, and make sure the core content is included. You can paraphrase if needed. Ask at most one question at a time.

The agenda is as follows:
1. Asking about the user's past two days of experience with respect to physical activity. Make sure you include all of the following questions in a natural manner:
a) How have you been these past two days?
b) How did your plan for physical activity go? Did you get a chance to [insert the activity previously discussed]?
c) When you were doing the physical activity, how did you feel emotionally?

2. Update the action plan for physical activity if needed.

3. Discuss the following specific topic with the user: DISCUSSION_TOPIC

You may find the following information useful:

RETRIEVAL_CONTENT

4. "On a scale of 0 - 5, how stressed are you feeling overall?" Once you ask this question, IMMEDIATELY respond with "FINISHED". No need to follow up.

**AVOID PROVIDING RESPONSES THAT ARE MULTIPLE PARAGRAPHS LONG.**

When you have finished the agenda, respond "FINISHED".
\end{lstlisting}

\section{Clinical Empathy Module: Technical Details}
\label{app:empathy_tech_details}

\subsection{Empathy Opportunity Classifier}
\label{app:emp_class}
\citet{rey2024digital} annotates ten patient long-form messages into different EOs, often including multiple labels per individual sentence. We first split all patient messages into individual sentences, which results in a total of 202 datapoints. Using DSPy \cite{khattab2023dspy}, we optimized a GPT-4o-mini-based classifier that takes as input the specific sentence, its surrounding contextual paragraph, and descriptions for each of the EOs, and outputs the top three possible EO predictions. We then deduplicate these predictions to produce the final set of predictions. Our optimization process aims to maximize the prediction accuracy. The optimized classifier was able to achieve a top-three classification accuracy of 85.3\%. We report our revised EO definitions in Table \ref{tab:all_eos}.

\begin{table}[ht]
\centering
\renewcommand{\arraystretch}{1.2}
\begin{tabular}{p{0.28\linewidth} p{0.34\linewidth} p{0.34\linewidth}}
\hline
\textbf{Empathy Opportunity Type} & \textbf{Definition} & \textbf{Examples} \\
\hline
negative\_feelings\_explicit &
EXPLICIT EXPRESSIONS OF NEGATIVE FEELINGS, such as an emotive behaviour or a mental state &
I am sad; I cried/shook/shouted... (emotive behavior); I felt alone/imagined the worst/thought I would die/remembered the pain/hated the pills (mental state); I cried when I found out \\
\hline
negative\_feelings\_implicit &
IMPLICIT EXPRESSION OF NEGATIVE FEELINGS through reference to a negative experience, such as fear, confusion, anxiety, or sadness &
I could not see any colours and everything was fuzzy (afraid/confused); I still had not received the letter from the hospital (anxious); I'm in the process of retiring... I will be 66 in February (anxious/sad); It's been 3 days and I haven't heard back from my GP \\
\hline
negative\_judgment\_explicit &
EXPLICIT EXPRESSION OF NEGATIVE JUDGEMENT (others or self) &
The nurse was very unhelpful; I am not a very strong person; She is such an irresponsible person \\
\hline
negative\_judgment\_implicit &
IMPLICIT EXPRESSION OF NEGATIVE JUDGEMENT (others or self) &
The nurse didn't see me straight away; I didn't cope very well when my father died; I could have done better \\
\hline
positive\_self\_judgment\_explicit &
EXPLICIT EXPRESSION OF POSITIVE SELF-JUDGEMENT &
I am a good father; I am an excellent runner \\
\hline
positive\_self\_judgment\_implicit &
IMPLICIT EXPRESSION OF POSITIVE SELF-JUDGEMENT &
I am eating healthier than ever!; I don't smoke anymore \\
\hline
negative\_appreciation\_explicit &
EXPLICIT EXPRESSION OF NEGATIVE APPRECIATION (things, events, actions) &
The dinner was so boring; The mouthwash is not helping me/is disgusting \\
\hline
negative\_appreciation\_implicit &
IMPLICIT EXPRESSION OF NEGATIVE APPRECIATION (things, events, actions) &
I don't notice any difference after taking the tablets; I am not sure this is something for me \\
\hline
general &
All other types of utterances. &
--- \\
\hline
\end{tabular}
\caption{The set of Empathy Opportunities used in our work.}
\label{tab:all_eos}
\end{table}

\subsection{Strategy Sampler}
\label{app:strat_sampler}
During inference time, we would perform weighted sampling for selecting specific empathetic response strategies that the chatbot would employ. The aggregated distribution for the final sampling step is formed by summing up the corresponding distributions of the top three predicted EOs from the classifier. We sample only one strategy if the EO classifier returns one classification (the empathy opportunity here is unambiguous), and two strategies otherwise.

The specific mappings between the EOs and the empathetic response strategies are as follows:

\begin{lstlisting}
{
    "negative_feelings_explicit": {
        "neutral_support_explicit_appreciation": 9,
        "sharing_feelings_views": 3,
        "understand_feelings_views": 4,
        "elicit_indirect_confirmation": 1,
        "elicit_indirect_coaching": 4,
        "elicit_indirect_feeling_invitation": 1,
        "elicit_indirect_feeling_negative": 1,
        "elicit_direct_feeling": 1,
    },
    "negative_feelings_implicit": {
        "acceptance_explicit_implicit_judgment": 11,
        "neutral_support_explicit_appreciation": 8,
        "neutral_support_normalization": 1,
        "sharing_feelings_views": 2,
        "understand_feelings_views": 9,
        "elicit_indirect_coaching": 14,
        "elicit_indirect_feeling_invitation": 3,
        "elicit_direct_feeling": 1,
    },
    "negative_judgment_explicit": {
        "acceptance_explicit_implicit_judgment": 2,
        "neutral_support_explicit_appreciation": 5,
        "understand_feelings_views": 3,
        "elicit_indirect_coaching": 3,
    },
    "negative_judgment_implicit": {
        "acceptance_explicit_implicit_judgment": 7,
        "acceptance_repetition": 1,
        "neutral_support_explicit_appreciation": 7,
        "sharing_feelings_views": 2,
        "understand_feelings_views": 3,
        "elicit_indirect_coaching": 15,
        "elicit_indirect_judgment": 1,
        "elicit_direct_feeling": 1,
        "elicit_feelings_3rd": 1,
    },
    "positive_self_judgment_explicit": {
        "acceptance_explicit_implicit_judgment": 2,
        "neutral_support_explicit_appreciation": 1,
        "sharing_feelings_views": 1,
        "elicit_direct_appreciation": 1,
        "elicit_indirect_coaching": 2,
        "elicit_direct_judgment": 1,
        "elicit_indirect_feeling_invitation": 1,
    },
    "positive_self_judgment_implicit": {
        "acceptance_explicit_implicit_judgment": 49,
        "acceptance_repetition": 1,
        "neutral_support_explicit_appreciation": 5,
        "neutral_support_explicit_judgment": 1,
        "neutral_support_normalization": 1,
        "sharing_feelings_views": 1,
        "elicit_indirect_confirmation": 3,
        "elicit_indirect_coaching": 7,
        "elicit_direct_feeling": 4,
    },
    "negative_appreciation_explicit": {
        "sharing_feelings_views": 3,
        "elicit_indirect_coaching": 9,
        "elicit_indirect_feeling_emotive": 1,
        "elicit_indirect_feeling_invitation": 1,
    },
    "negative_appreciation_implicit": {
        "acceptance_explicit_implicit_judgment": 2,
        "acceptance_repetition": 1,
        "neutral_support_explicit_appreciation": 10,
        "sharing_feelings_views": 11,
        "understand_feelings_views": 10,
        "elicit_indirect_coaching": 11,
        "elicit_indirect_feeling_invitation": 3,
    },
    "general": {
        "acceptance_repetition": 7,
        "neutral_support_explicit_appreciation": 1,
        "neutral_support_explicit_judgment": 1,
        "sharing_feelings_views": 3,
        "elicit_indirect_coaching": 1,
        "elicit_indirect_feeling_invitation": 2,
    },
}

\end{lstlisting}

When the strategies are sampled, their corresponding definitions are concatenated and incorporated into a system prompt so that our chatbot can follow the specific strategies when responding. We report the definitions for each of the empathetic response strategies in Tables \ref{tab:ers-definitions-1} and \ref{tab:ers-definitions-2}.



\begin{table}[ht]
\centering
\renewcommand{\arraystretch}{1.15}
\begin{tabular}{p{0.28\linewidth} p{0.68\linewidth}}
\toprule
\textbf{Empathetic Response Strategy} & \textbf{Definition} \\
\midrule
acceptance\_explicit\_implicit\\\_judgment &
Expressing acceptance in response to user's explicit, implicit or potential negative or positive self-judgement; Unconditional positive regard (or ``praise'') through: Explicit expression of positive judgement of the user as a person (``You are a fantastic mom''); Implicit expression of positive judgement through explicit positive appreciation of the user's actions or thoughts (``It looks like you are making great progress'') \\
\midrule
acceptance\_repetition &
Unconditional positive regard (or ``praise'') through repetition or paraphrasing of user's words and avoidance of immediate countering statements or premature reassurance \\
\midrule
neutral\_support\_explicit\\\_appreciation &
Neutral support (even when approval cannot be granted, withholding judgement): Explicit appreciation of the user's behaviour, ideas or feelings in terms of their ``normality'' and ``acceptability'' (``It is (completely) normal/not unusual/acceptable\dots to do/think/feel X'', ``It is not (at all) surprising/crazy to do/think/feel X'') \\
\midrule
neutral\_support\_explicit\_judgment &
Neutral support (even when approval cannot be granted, withholding judgement): Explicit expressions of judgement when denying potential negative self-assessment by the user (``You are not odd, bad, crazy\dots for doing/thinking/feeling X\dots'') \\
\midrule
neutral\_support\_normalization &
Neutral support (even when approval cannot be granted, withholding judgement): Implicit appreciation of the user's behaviour, ideas or feelings in terms of their ``normality'' and ``acceptability'' \\
\midrule
sharing\_feelings\_views &
Sharing the user's feelings or views through expressions of agreement (``emotive empathy''); Shared feelings (``I would also\dots\ if I were you''), shared feelings through interjection and intonation (``Oh no!''), shared judgement (``Yes, your boss could have been more understanding''), shared appreciation (``Yes, this is a difficult exercise'') \\
\midrule
understand\_feelings\_views &
Expressing explicit understanding or acknowledgement of users' feelings and views; Formulations including verbs of acknowledgement (``I understand/see/realise/appreciate\dots''), adjectival constructions expressing understanding (``It is clear/apparent\dots\ to me''), or alternative formulations (``It strikes me that\dots'', ``I am aware/conscious that\dots''). \\
\midrule
elicit\_direct\_appreciation &
Direct elicitation of user's appreciation, e.g.\ ``How would you rate your progress so far?'' \\
\midrule
elicit\_indirect\_confirmation &
Indirect elicitation of user's appreciation through: Invitation of user's confirmation, rejection or clarification of interpreted view of things, events and actions (``You do not seem to find that book so useful'') \\
\bottomrule
\end{tabular}
\caption{Empathetic response strategies and definitions (part 1 of 2).}
\label{tab:ers-definitions-1}
\end{table}

\begin{table}[ht]
\centering
\renewcommand{\arraystretch}{1.15}
\begin{tabular}{p{0.28\linewidth} p{0.68\linewidth}}
\toprule
\textbf{Empathetic Response Strategy} & \textbf{Definition} \\
\midrule
elicit\_indirect\_coaching &
Indirect elicitation of user's appreciation through: Enquires about the physical activity counseling (``Would you like us to review your goals together?'') \\
\midrule
elicit\_direct\_judgment &
Direct elicitation of user's judgement, e.g.\ ``How are you finding your coach?'', ``Was the nurse helpful?'', ``How committed are you to improving your health?'' \\
\midrule
elicit\_indirect\_judgment &
Indirect elicitation of user's judgement through: Enquiries about others or user's behaviour (``How did your husband react?''); Invitation of user's confirmation, rejection or clarification of interpreted views of others or themselves (``You seem very motivated to start your health programme'') \\
\midrule
elicit\_indirect\_feeling\_emotive &
Indirect elicitation of user's feelings through: Enquiries about the ``emotive behaviour'' displayed (``Why are you crying?'') \\
\midrule
elicit\_indirect\_feeling\_invitation &
Indirect elicitation of user's feelings through: Invitation of user's confirmation, rejection or clarification of interpreted affectual states (``You seemed concerned when I mentioned diabetes'') \\
\midrule
elicit\_indirect\_feeling\_negative &
Indirect elicitation of user's feelings through: Questions about potentially negative/critical experiences (``Did you have a happy childhood?'') \\
\midrule
elicit\_direct\_feeling &
Direct elicitation of user's feelings; Enquiries about mental state (``How/what do/did you feel/think/expect?'') or emotive behaviour (``How did/do you react?'') \\
\midrule
elicit\_feelings\_3rd &
Indirect elicitation of user's feelings and views through formulations in which feelings and evaluations (potential or real) are attributed to third parties who might find/found themselves in circumstances similar to the user's; ``Many people in your position would be quite annoyed [affect]/would find this annoying [appreciation].'', ``My sister went through something similar and she struggled to see the benefits of [affect]/to value [appreciation] that opportunity'' \\
\bottomrule
\end{tabular}
\caption{Empathetic response strategies and definitions (part 2 of 2).}
\label{tab:ers-definitions-2}
\end{table}

\section{Pilot Study}
\label{app:pilot_study}

We conducted a between-subject pilot study through Prolific to explore whether different versions of the chatbot would induce short-term effects on self-efficacy in participants, in addition to ensuring the workflow is functional as intended. For this pilot study, we focused primarily on examining the Empathetic and Non-Empathetic versions of the chatbot, with nine participants for Empathetic, eight participants for Non-Empathetic, and four participants for Standard. Participants were paid \$8 for 30 minutes of their time. 

\subsection{Study Design}
Upon signing the informed consent, participants were directed to fill out a pre-survey about participant demographic information and their physical activity levels. Then, participants engaged with the different versions of the chatbot in a short-form conversation setting. Specifically, they completed the initial session with the chatbot, and are pinged to complete a second session (maintenance session) a few minutes after their first session. They are instructed to pretend that two days have passed, and that they failed to complete their physical activity goals as set in the initial session. We hypothesize that the contrast between the Non-Empathetic and Empathetic chatbot versions would be more prominent when the user performs negative self-disclosure. After the two sessions, participants filled out a survey discussing their conversational experience.

We use conversations collected in this pilot study to validate that the exhibited empathy levels in the different chatbot versions are appropriate, both through LLM judgment and human evaluation. 

\subsection{Participant Characteristics}

We report the statistics of the participants in Table \ref{tab:pilot_study_participants}. For reporting PA levels, participants indicated their current stage of behavioral change through answering the following question:

\noindent \textbf{Question:} Please select the one that describes your current physical activity status.
\begin{enumerate}[label=(\Alph*)]
    \item I am currently not physically active and do not intend to engage in physical activity in the next six months
    \item I am currently not physically active, but I am thinking about getting more physically active in the next six months
    \item I currently do some physical activity, but not regularly
    \item I am currently physically active, but have only begun doing so within the last six months
    \item I am currently physically active and have done so for more than six months
\end{enumerate}

\begin{table}[!h]
    \centering
    \begin{tabular}{c|cccccc}
    \toprule
       \textbf{Condition} & \textbf{Age} & \textbf{Gender} & \textbf{Step Count} & \textbf{PA Status}\\
       \midrule
       NE & 42 & Male & N/A & (E) \\
       NE & 50 & Male & N/A & (E)\\
       NE & 66 & Female & N/A & (E)  \\
       NE & 34 & Female & N/A & (C) \\
       NE & 54 & Female & N/A & (C)\\
       NE & 33 & Male & N/A & (E)  \\
       NE & 20 & Female & N/A & (B) \\
       NE & 55 & Male & N/A & (B)  \\
       \midrule
       S & 54 & Male & 10000 & (C)  \\
       S & 27 & Male & Unknown & (E)  \\
       S & 26 & Male & 6000 & (E)\\
       S & 40 & Male & 12000 & (E)  \\
       \midrule
       E & 25 & Male & 4000 & (D) \\
       E & 21 & Male & 600 & (B) \\
       E & 25 & Male & 5000 & (C) \\
       E & 49 & Male & 1200 & (C) \\
       E & 28 & Male & Unknown & (D) \\
       E & 36 & Male & 6000 & (C)  \\
       E & 25 & Male & 6000 & (D) \\
       E & 34 & Male & 7500 & (E) \\
       E & 31 & Male & 100 & (B)\\
       \bottomrule
    \end{tabular}
    \caption{Participant statistics for the pilot study. The conditions are Non-Empathetic (NE), Standard (S), and Empathetic (E). The PA Status column refers to the choice made by the participant with respect to their stage of behavioral change.}
    \label{tab:pilot_study_participants}
\end{table}

We notice that the majority of our participants are currently physically active, which may lead to a lower evaluation of chatbot suggestion usefulness. As Prolific has not rolled out the ability to screen out participants during the time of the pilot study, we were not able to sufficiently filter out these active participants. As a result, we primarily focus on assessing the usability levels of the different chatbot versions, as well as free-form feedback provided by the participants. 

\subsection{Results and Analysis}

\subsubsection{Usability Results}
We document the raw counts of Likert ratings due to the small sample sizes (Table \ref{tab:usability_pilot}). To our surprise, we find that the Non-Empathetic chatbot is consistently better rated in our pilot study compared to the empathetic version. In particular, according to the median ratings, participants find the Non-Empathetic chatbot to be more enjoyable and more satisfying, while the Empathetic chatbot is easier to use. 

\begin{table}[!h]
    \centering
    \begin{tabular}{c|c|ccccc|cc}
    \toprule
       \textbf{Question} & \textbf{Condition} & \textbf{\# Score = 1} & \textbf{\# Score = 2}  & \textbf{\# Score = 3} & \textbf{\# Score = 4} & \textbf{\# Score = 5} & \textbf{Mean} & \textbf{Median} \\
       \midrule
       Ease of Use & NE & 4 & 2 & 0 & 1 & 1 & 2.13 & 1.5 \\
        & S & 2 & 1 & 1 & 0 & 0 & 1.75 & 1.5 \\
        & E & 6 & 2 & 1 & 0 & 0 & 1.44 & 1.0 \\
        \midrule
        Enjoyment & NE & 0 & 1 & 0 & 2 & 5 & 4.38 & 5.0 \\
         & S & 0 & 0 & 1 & 1 & 2 & 4.25 & 4.5 \\
         & E & 0 & 1 & 2 & 3 & 3 & 3.89 & 4.0 \\
        \midrule
       Satisfaction & NE & 0 & 0 & 1 & 3 & 4 & 4.38 & 4.5 \\
        & S & 0 & 0 & 1 & 2 & 1 & 4.00 & 4.0 \\
        & E & 0 & 1 & 2 & 3 & 3 & 3.89 & 4.0 \\
       \midrule
       Time & NE & 0 & 0 & 1 & 2 & 5 & 4.50 & 5.0 \\
       & S & 0 & 0 & 0 & 0 & 4 & 5.00 & 5.0 \\
       & E & 0 & 0 & 3 & 1 & 5 & 4.22 & 5.0 \\
       \midrule
       Understandable & NE & 0 & 0 & 0 & 2 & 5 & 4.50 & 5.0 \\
       & S & 0 & 0 & 0 & 0 & 4 & 5.00 & 5.0 \\
       & E & 0 & 0 & 1 & 2 & 6 & 4.56 & 5.0 \\
    \bottomrule
    \end{tabular}
    \caption{The raw counts of the ratings for each of the chatbot usability questions, in addition to the means and averages, for each of the empathy conditions.}
    \label{tab:usability_pilot}
\end{table}

\subsubsection{Conversation Quality Results}
Participants rated the naturalness of responses similarly between the three conditions, with median ratings of 4.5 for NE, 4.5 for S, and 4 for E. The slightly lower rating of naturalness from the Empathetic condition can be attributed to the generally longer responses from the chatbot and to how the Empathetic chatbot is designed to express empathy with every utterance.

When it comes to features of the conversations that they liked, five participants across all three conditions responded with the stress relief component, indicating that stress relief can be a useful tool for enhancing user engagement. Another common theme across conditions is personalization; some participants generally enjoyed that the suggestions are tailored to their specific situations, whereas other participants wished that the advice was less vague and even more personalized. 

We notice that chatbot perception is influenced by significant individual variations. All three versions of the chatbot can be perceived as friendly and even empathetic. However, participants place emphasis differently when describing these versions. Both the Empathetic and the Standard chatbots received the descriptor ``like chatting with a friend'', while the Empathetic chatbot is praised for its positive energy:

\begin{quote}
    \textit{I also loved how it was genuinely happy for my success in achieving my goals.}

    \textit{[I]t’s almost like a buddy}

    \textit{I enjoyed the reassurance and positive energy it gave me!}

    \textit{[I]t truly cares about your wellbeing.}
\end{quote}

Notably, although some participants deem the Non-empathetic chatbot to be ``sympathetic and down-to-earth'', one of the participants said that the Non-empathetic conversation ``felt very rote'' and that ``there was no standout moment''. More particularly:

\begin{quote}
    \textit{I do not feel as if the responses to my input were expanded upon in any meaningful way. }
\end{quote}

With respect to areas of improvement, the core feedback from participants includes repetition within session, technical issues, and overly long responses. We aimed to address these issues in our chatbots before running the official user study.




\subsubsection{Overall Discussion}

The results of our pilot study call into question our assumption that users always prefer empathetic physical activity coaching chatbots. These preliminary findings additionally suggest that people possess different capabilities in detecting empathy, and may have varied individual preferences in terms of empathy levels in a physical activity coaching chatbot. Informed by this, we add questions in our pre-survey for the official user study about subjective preferences for how empathetic a coaching chatbot should be, as well as subjective preferences for the extent of personalization. 

Our pilot study further demonstrates the potential effect of individual differences on our experimental results. Due to this variance caused by individuality, we decided to employ a within-subject design to better compare the effect of different empathetic conditions on participants.

It is, however, still important to highlight that the results of the pilot study may be related to the age and the general level of comfort with chatbots among our participants. Furthermore, participants may artificially rate the chatbots highly across conditions due to the desire to please the experimenter, and because they may generally expend less effort.

\section{Official User Study: Participant Familiarity with LLMs}
\label{app:llm_prefs_participants}

After the study, we surveyed our participants for their familiarity with LLMs (``How often do you use LLM-based chatbots (e.g. ChatGPT, Claude, Gemini)?'') and their general attitude toward Generative AI (``What is your general attitude toward Generative AI?'', 5-point scale, 1 = very negative, 5 = very positive). Unfortunately, only 10 participants responded. We record the results below:

\begin{table}[!h]
    \centering
    \begin{tabular}{c|cc}
    \toprule
       \textbf{Experiment ID} & \textbf{Frequency of LLM Use} & \textbf{General Attitude} \\
       \midrule
       P3 & A few times a week & 4 \\
       P1 & Every day & 4 \\
       P4 & Every day & 4 \\
       P12 & Never & 2 \\
       P9 & Every day & 3 \\
       P13 & Every day & 5 \\
       P7 & A few times a week & 4 \\
       P2 & A few times a month & 2 \\
       P11 & A few times a week & 3 \\
       P8 & A few times a week & 4 \\
    \bottomrule
    \end{tabular}
    \caption{Participant familiarity with LLM chatbots and their attitude towards Generative AI.}
    \label{tab:placeholder}
\end{table}

We see that, in general, our participants use LLM chatbots fairly frequently, and have positive attitudes towards Generative AI.

\section{Check-in Form Results: Further Analysis of Change Over Time}
\label{app:check_in_form_change}

We model the number of days since enrollment as the independent variable, and the self-report measures (step counts, intention to follow, self-efficacy, and chatbot usefulness) as the dependent variables. 

The linear regressions demonstrate positive relationships between the number of days and all self-reported measures ($\beta_{\text{intention}} = 0.003, \beta_{\text{efficacy}} = 0.005, \beta_{\text{usefulness}} = 0.002$), although only the increase in step counts is statistically significant ($\beta_{\text{steps}} = 67.5, R^2= 0.05, p=0.0004$). We observe a similar trend in the average and mean weekly changes from enrollment to the end of engagement, with a positive median only for step counts (Table \ref{tab:deltas_avg_means}).

\begin{table}[!h]
    \centering
    \begin{tabular}{l|cc}
    \toprule
    \textbf{Form Item} & \textbf{Mean} & \textbf{Median}\\
    \midrule
    $\Delta$ Steps & 674.1 $\pm$ 359.9 & 500.0 \\
    $\Delta$ Intention & -0.10 $\pm$ 0.12 & 0.0 \\
    $\Delta$ Self-Efficacy & -0.03 $\pm$ 0.14 & 0.0 \\
    $\Delta$ Usefulness & -0.15 $\pm$ 0.26 & 0.0 \\
    \bottomrule
    \end{tabular}
    \caption{The amount of \textbf{change per week} for each self-report measure in the check-in form, across participants.}
    \label{tab:deltas_avg_means}
\end{table}

Participants generally experience improvements in step counts throughout the user study. We discuss a possible reason for this in Section \ref{sec:user_interviews}. However, participating in the study does not seem to have a significant effect on the subjective measures in our check-in form, plausibly due to changes in the empathetic condition.

\section{User Exit Interview Questionnaire}
\label{app:user_exit_interview_questions}

\begin{enumerate}
    \item How was your general experience using the chatbot?
    \begin{enumerate}
        \item Was the chatbot generally fluent? General impression across the weeks?
    \end{enumerate}
    \item Were there specific anecdotes that you would like to share?
    \item Can you feel that the different versions of the chatbot were different between weeks? 
    \begin{enumerate}
        \item How could you tell? 
        \item How did that affect you? Reduce physical activity barriers? Increase motivation? Increase step count?
    \end{enumerate}
    \item How can we improve the chatbot?
    \item How can the chatbot be more helpful to you in terms of improving your physical activity?
    \item Was the chatbot helpful to you in reducing your stress levels?
    \begin{enumerate}
        \item Were you able to do more physical activity after stress reduction?
    \end{enumerate}
    \item Do you trust the advice given by the chatbot?
    \begin{enumerate}
        \item Do you feel like it would be different had it been given by a human?
    \end{enumerate}
    \item Do you feel that the chatbot advice is feasible?
\end{enumerate}
\end{document}